\documentclass[aps,showpacs]{revtex4}
\usepackage{epsfig}
\usepackage{natbib}
\usepackage{amsmath}
\usepackage{times}
\usepackage{psfrag}
\usepackage{subfigure}
\begin{document}

\title{
Global stability analysis of birhythmicity in a self-sustained
oscillator}
\author{R. Yamapi}
\email[]{Corresponding author: Email: ryamapi@yahoo.fr}
\affiliation{Fundamental Physic's Laboratory, Group of Nonlinear
Physics and Complex System,Department of Physics, Faculty of
Science, University of Douala,  Box 24 157 Douala, CAMEROON}

\author{G. Filatrella}
 \affiliation{Laboratorio Regionale SuperMat,
INFM/CNR Salerno and Dipartimento di Scienze Biologiche ed
Ambientali, Universit\`a del Sannio, via Port'Arsa 11, I-82100
Benevento, ITALY}

\author{M. A. Aziz-Alaoui }
 \affiliation{  Applied Mathematics Laboratory, University of Le Havre,
  25 rue ph. Lebon, B.P
 540, Le Havre, Cedex, FRANCE}
\date{July 13, 2009}

\begin{abstract}

We analyze global stability properties of birhythmicity in a
self-sustained system with random excitations. The model is a
multi-limit cycles variation of the  van der Pol oscillator
introduced to analyze enzymatic substrate reactions in brain
waves. We show that the two frequencies are strongly influenced by
the nonlinear coefficients $\alpha$ and $\beta$. With a random
excitation, such as a Gaussian white noise, the attractor's global
stability is measured by the mean escape time $\tau$ from one
limit-cycle. An effective activation energy barrier is obtained by
the slope of the linear part of the variation of the escape time
$\tau$ versus the inverse noise-intensity $1/D$. We find that the
trapping  barriers of the two frequencies can be very different,
thus leaving the system on the same attractor for an overwhelming
time. However, we also find that the system is nearly symmetric in
a narrow range of the parameters.
\end{abstract}
 \maketitle

{\bf

Some models employed to describe natural systems, such as for
instance glycolysis reactions and circadian proteins rhythmics,
exhibit spontaneous oscillations at two distinct frequencies. The
phenomenon is known as birhythmicity, and the underlying dynamical
structure is characterized by the coexistence of two stable
attractors, each displaying a different frequency. Being the
attractors locally stable, the system would however stay at a
single frequency, the one selected by the choice of the initial
conditions, unless an external source disturbs the evolution and
causes a switch to the other attractor. To investigate such
process, we have focused on a particular system of biological
interest, a modified van der Pol oscillator (that displays
birhythmicity), to determine  the global stability properties of
the attractors under the influence of noise. More specifically, we
have characterized the stability of the attractors with the escape
times, or the average time that the system requires to switch from
an attractor to the other under the influence of random
fluctuations. Such analysis reveals that the two attractors can
possess very different properties, with very different relative
residence times. Even excluding the most asymmetric cases, the
system can spend something like 10 years on one attractor for each
second spent on the other. We conclude that although a system can
be structurally biorhythmic for the contemporary presence of two
locally stable attractors at two different frequencies, actual
switch from one frequency to the other could be very difficult to
observe. A global stability analysis can therefore help to
determine the region of the parameter space in which birhythmic
behavior will be genuinely observed.

}

\newpage

\section{Introduction}
\noindent


Self-oscillating systems exhibit limit cycles, or periodic
sustained oscillations. Examples are abundant, with periods
ranging from cardiac rhythms of seconds, glycolysis over the
minutes, circadian oscillations over the 24 hours, while
epidemiological oscillations extend even over the years
\cite{goldbeter-1996,winfree,goldbeter-2002}. Birhythmicity refers
to the coexistence of two attractors characterized by two
different amplitudes and two frequencies: depending on the initial
conditions, the system can produce self--oscillations at two
distinct periods. Such hysteretic behavior has been sometimes
observed in biological systems \cite{bistabexp}. Many more
theoretical studies have shown the possible occurrence of
birhythmicity in models of glycolytic oscillations
\cite{decroly-goldbeter}, chemical kinetic equations
\cite{morita}, circadian proteins rhythmics
\cite{leloup,stich,tsumoto}, and biochemical reactions
\cite{fuente}. Perhaps the simplest model that exhibits
birhythmicity is a variation of the well known van der Pol
oscillator proposed by Kaiser \cite{kaiser1} to model enzyme
reactions. In such a model it has been shown that two attractors
can coexist for some values of the parameters
\cite{kaiser1,frohlich,enjieu-chabi-yamapi-woafo}, and
birhythmicity is robust enough to enable two
\cite{enjieu-yamapi-chabi} or more \cite{yamapi-enjieu-filatrella}
oscillators to synchronize. The aim of this work is to adopt the
Kaiser modification of the van der Pol oscillator
\cite{kaiser1,kaiserv,kaiserp} as a paradigm for birhythmicity to
analyze the global stability properties of the attractors under
the influence of random excitations, i.e. the response to finite
perturbations \cite{kautz,dykman78,dykman01}. In fact while local
stability properties that refer to small perturbations of the
steady state have been analyzed in
Ref.\cite{yamapi-enjieu-filatrella}, global stability refers to
the response to large random fluctuations (large enough to drive
the system from one attractor to the other). Such global stability
property has not been addressed for the model proposed in Ref.
\cite{kaiser1,frohlich,enjieu-chabi-yamapi-woafo}, and seldom
investigated in birhythmic systems (see Ref. \cite{kar-ray} for an
exception). Global stability is well studied in ac driven (and
hence monorhythmical) systems
\cite{dykman78,dykman01,stambaugh02,stambaugh06}, for instance in
connection with the phenomenon of stochastic resonance
\cite{almog07} or of switching between chaotic attractors
\cite{luchinsky02,kraut04}. We want here to focus on the passage
between two attractors characterized by two different frequencies,
and therefore we will emphasize the consequences of noise driven
switching on the birhythmic properties, while in periodically
driven systems the frequency is pre-selected by the external
drive.

When noise is added, the mean time $\tau$ required to escape from
a basin of attraction is a useful measure of the attractor's
global stability also for non equilibrium or oscillating systems,
such as ac-driven Josephson circuits with intrinsic thermal
fluctuations \cite{kautz} or with finite-spectral-linewidth ac
current \cite{GF}. In the same spirit, we propose to measure the
attractor's global stability with the mean escape time $\tau$ from
one stable limit-cycle attractor to another stable limit cycle
attractor. Escape occurs when, under the influence of a
deterministic or random term, the system crosses the boundary of
the basin of attraction ({\it i.e.} it is driven across the
unstable limit-cycle).

Let us remark that even if we focus on switches due to random
perturbations, one could also drive the system from an attractor
to the other by means of a deterministic or structural change.
This type of switch will be not considered in the present work,
 however it is also possible from the deterministic dynamics -- considering all possible paths that lead from one attractor to the other with the appropriate noise-dependent weight -- to retrieve the escape rate \cite{graham81,freidlin,graham85,kautz,dykman01}.

We will show that the reason that might hamper actual observation
of birhythmicity in a noisy environment is the asymmetry of the
escape times. In such a case the system is likely to stay for a
much longer time on one attractor with respect to the other, and
therefore one would rarely observe the spontaneous transition from
an attractor to the other \cite{dykman78,dykman01,stambaugh06}. We
conclude that although coexistence of two stable attractors with
different frequencies is a prerequisite for birhythmicity, actual
observation might be hindered by very asymmetric stability
properties of the two attractors. In other words we will consider
birhythmical systems as bistable systems and the numerically
evaluated escape times will serve as a measure of the relative
stability of the two solutions. For a glycolytic model it has
indeed been proven by means of the Fokker-Planck equation
associated to the weak noise limit  that the original system with
two stable attractors (and hence with birhythmical behavior)
changes structures and becomes monorhythmical \cite{kar-ray}. Our
analysis arrives at a similar conclusion: the escape time from one
of the attractors might be very large compared to the escape time
of the reverse process, even by many orders of magnitude. In
addition, we find that for some range of parameters the system is
(approximately) symmetric. In this (indeed narrow) parameter space
region the two attractors have comparable properties, and
birhythmicity is more likely to be observed.

The paper is organized as follows. In section {\it II}, we
describe the self-sustained system with random excitation and the
algorithm of the numerical simulations. Section {\it III} deals
with the dynamical attractors of free-noise multi-limit-cycles
self-sustained system. We will show that birhythmicity features
are not uniform in the parameter region where it appears in the
modified  van der Pol system. In section {\it IV}, we focus on
numerical computed escape rates using the Box-Mueller random
Gaussian generator algorithm \cite{knuth} for numerical
integration with the Euler method. The Arrhenius factor ({\it
i.e.} the relation  between the escape time $\tau$,
 and the noise intensity $D$), allows us to determine an effective
activation energy barrier $\Delta U_i$, or the slope of the linear
part of the variation of the escape time versus the inverse
noise-intensity, as a useful method to summarize the results. The
last section is devoted to conclusions.

\section{The self-sustained system with random excitation}
\subsection{The multi-limit cycle van der Pol oscillator}
\noindent

The model considered is a van der Pol-like oscillator with a
nonlinear function of higher polynomial order described by the
following nonlinear equation (overdots as usual stand for the
derivative with respect to time)
\begin{eqnarray}
\label{eq1} \ddot x-\mu (1- x^2+\alpha x^4-\beta x^6)\dot x+x=0,
\end{eqnarray}
where $\alpha,\beta$ and $\mu$ are positive parameters that tune
the nonlinearity. Model (1) is therefore a prototype for
self-sustained systems and exhibits some interesting features of
nonlinear dynamical systems; for instance Ref.
\cite{kaiserv,kaiserp} have analyzed the super-harmonic resonance
structure and have found symmetry-breaking crisis and
intermittence. The nonlinear dynamics and the synchronization
process of two such systems have been recently investigated in
Ref.\cite{enjieu-chabi-yamapi-woafo,enjieu-yamapi-chabi}, while
the possibility that introducing an active control of chaos can be
tamed for an appropriate choice of the coupling parameters has
been considered in Ref. \cite{yamapi-nana-enjieu-2007}.

Eq.~(\ref{eq1}) describes several dynamic systems, ranging from
physics to engineering and biochemistry \cite{vanderpolgeneral}.
In particular Eq.~(\ref{eq1}) seems to be more appropriate for
some biological processes than the classical van der Pol
oscillator, as shown by Kaiser in Ref. \cite{kaiser2}. When
employed to model biochemical systems, namely the
enzymatic-substrate reactions, $x$ in Eq.~(\ref{eq1}) is
proportional to the population of enzyme molecules in the excited
polar state, the quantities $\alpha$ and $\beta$ measure the
degree of tendency of the system to a ferroelectric instability,
while $\mu$ is a positive parameter that tunes nonlinearity
\cite{enjieu-chabi-yamapi-woafo}.

The nonlinear self-sustained oscillator Eq.~(\ref{eq1}) possesses
more than one stable limit-cycle solution \cite{kaiser2}, a
condition for the occurrence of birhythmicity. Birhythmic systems
are of interest, for example in biology, to describe the
coexistence of two stable oscillatory states, a situation that can
be found in some enzyme reactions \cite{li}. Another example is
the explanation of the existence of multiple frequency and
intensity windows in the reaction of biological systems when they
are irradiated with very weak electromagnetic fields
\cite{kaiserp,kaiser2,kaiser3,kaiser4,kaiser5,kaiser6}.  In this
work we will focus on model (\ref{eq1}) as a prototype for the
occurrence of birhythmicity.

\subsection{The model with random excitation and algorithm for numerical simulations}
\noindent

Let us consider the multi-limit-cycle van del Pol-like oscillator
Eq.~(\ref{eq1}) to model coherent oscillations in biological
systems, such as an enzymatic substrate reaction with
ferroelectric behavior in brain waves models (see
Ref.\cite{kaiser1,frohlich,enjieu-chabi-yamapi-woafo} for more
details). In this case, one should include the electrical field
applied to the excited enzymes, which depends for example on the
external chemical influences ({\it i.e.}, the flow of enzyme
molecules through the transport phenomena). One can therefore
assume that the external chemical influence contains a random
perturbation. Therefore, adding both the chemical and the
dielectric contribution, the activated enzymes are subject to a
random excitation governed by the Langevin version of
Eq.~(\ref{eq1}), namely:
\begin{eqnarray}
\label{eq2} \ddot x-\mu (1- x^2+\alpha x^4-\beta x^6)\dot
x+x=\Gamma (t),
\end{eqnarray}
where $\Gamma(t)$ is a Gaussian additive white noise
\cite{middleton} whose statistical features are completely
determined by the additional properties:
\begin{eqnarray}
\label{eq3} &&<\Gamma (t)>=0\nonumber \\
&&<\Gamma (t) \Gamma (t')>=2D\delta (t-t').
\end{eqnarray}
The white-noise quality of $\Gamma$ is contained in the Dirac
$\delta$-function correlation (\ref{eq3}). The parameter $D$ is
the intensity of the Gaussian white noise.

In this work we will numerically integrate Eqs.
(\ref{eq2},\ref{eq3}) using a Box-Mueller algorithm \cite{knuth}
to generate the Gaussian white noise from two random numbers $a$
and $b$ which are uniformly distributed on the unit interval
$[0,1]$. By introducing the new variable $\dot x=u$, Eq.
(\ref{eq2}) can be written in the form
\begin{eqnarray}
\label{eqmodel}
\dot x&=&u  \hspace{14.5cm} (4a) \nonumber \\
\dot u&=&\mu(1-x^2+\alpha x^4-\beta x^6)u-x +\Gamma \hspace{9cm}
(4b) \nonumber \setcounter{equation}{5}
\end{eqnarray}

\noindent The simple Euler algorithm version of the integration of
equation (\ref{eqmodel}) is given by
\begin{eqnarray}
\label{euler}
&&\Gamma_{\Delta t}=\sqrt{-4D\Delta t \, log(a)}\cos (2\pi b),\hspace{9.4cm} (5a)\nonumber \\
&&x|_{t+\Delta t}=x+u\Delta t,\hspace{12cm} (5b)  \nonumber\\
 &&u|_{t+\Delta
t}=u+(\mu(1-x^2+\alpha x^4-\beta x^6)u-x)\Delta t+\Gamma_{\Delta
t}.\hspace{5.9cm} (5c)\nonumber \setcounter{equation}{6}
\end{eqnarray}

 The step size used for numerical
integration is generally equal to $\Delta t=0.0001$, but in some
cases we have used a smaller step. We have also checked that
averaging over as many as $200$ realizations the results converge
within few percents. We notice that there are more accurate
methods to estimate the escape from a basin of attraction, or in
general close to an absorbing barrier, to avoid the inaccuracy due
to a finite sampling of the random evolution \cite{mannella}.
However, we have carefully checked that the results we have
obtained are independent of the step size. This has been done in
two ways: halving the step size until stable results are reached
(and with much attention to low noise intensity
$D$\cite{mannella}) and  calibrating the numerical method with a
potential with a well defined activation barrier to retrieve the
Kramer escape rate \cite{kramer}.

So, although analytical treatments based on the Fokker-Planck
version of the Langevin equation (\ref{eq2}) \cite{tutorial}, the
variational approach
\cite{graham81,freidlin,graham85,kautz,dykman01}, or faster
numerical algorithms such as the stochastic version of the
Runge-Kutta methods are available, we have preferred to use the
simple procedure given by Eq.~(\ref{euler}) that proved fast
enough for the present project.

{\scriptsize
\begin{eqnarray*}
\begin{tabular}{|l|c|c|c|l|}
\hline
$S_i=(\alpha,\beta)$& Analytical Amplitude & Numerical Amplitude & Analytical Frequency & Numerical Frequency \\
\hline
                   &$A_1$=2.37720          & $A_1$=2.378          & $\Omega_1$=1.00212      & $\Omega_1$=1.00015 \\

$S_1=(0.114;0.003)$& $A_2$=5.02638           & Unstable            & $\Omega_2$=1.00113       & Unstable         \\

                 & $A_3$=5.46665          & $A_3$=5.464         & $\Omega_3$=1.0231      & $\Omega_3$=1.019575 \\
\hline
                        & $A_1$=2.3069          & $A_1$=2.30265          & $\Omega_1$=0.987      & $\Omega_1$=0.988 \\

$S_2=(0.1;0.002)$     & $A_2$=4.8472           & Unstable            & $\Omega_2$=1.000113       & Unstable         \\

                        & $A_3$=7.1541         & $A_3$=7.1345         & $\Omega_3$=0.97123      & $\Omega_3$=0.97831 \\
\hline
                        & $A_1$=2.4269         & $A_1$=2.4259          & $\Omega_1$=0.985      & $\Omega_1$=0.988 \\

$S_3=(0.12;0.003)$     & $A_2$=4.2556        & Unstable            & $\Omega_2$=0.999       & Unstable         \\

                        & $A_3$=6.3245            & $A_3$=6.33918         & $\Omega_3$=0.9865      & $\Omega_3$=0.988 \\
\hline
                        & $A_1$=2.4903          & $A_1$=2.48971         & $\Omega_1$=1.000212      & $\Omega_1$=1.000507 \\

$S_4=(0.13;0.004)$     & $A_2$=4.4721           & Unstable            & $\Omega_2$=1.000113       & Unstable         \\

                        & $A_3$=5.0791         & $A_3$=5.07739         & $\Omega_3$=0.99912      & $\Omega_3$=0.9989 \\
\hline
                         & $A_1$=2.6605         & $A_1$=2.65963          & $\Omega_1$=1.000212      & $\Omega_1$=1.000507 \\

$S_5=(0.145;0.005)$    & $A_2$=3.8305           & Unstable            & $\Omega_2$=1.000113       & Unstable         \\

                        & $A_3$=4.964           & $A_3$=4.96336         & $\Omega_3$=1.00049903      & $\Omega_3$=1.000256 \\
\hline
                        & $A_1$=2.7864          & $A_1$=2.78532         & $\Omega_1$=0.99923     & $\Omega_1$=0.9989 \\

$S_6=(0.154;0.006)$     & $A_2$=3.8821           & Unstable            & $\Omega_2$=1.000113       & Unstable         \\

                        & $A_3$=4.2698         & $A_3$=4.26807         & $\Omega_3$=1.000231      & $\Omega_3$=1.000507 \\
\hline
\end{tabular}
\end{eqnarray*}}
{\it Table 1: \small Comparison between analytical and numerical
characteristics of the limit cycles. All data refer to the case $\mu=0.1$.}\\

In the absence of noise ($\Gamma=0$), Eq.~(\ref{eq2}) reduces to
the modified version of the  van der Pol oscillator (see
Eq.~(\ref{eq1})), which has steady-state solutions that correspond
to attractors in state space and depend on the parameters $\alpha,
\beta$ and $\mu$. Before taking up the subject of noise-induced
transitions between dynamical attractors, we focus in the
following section on the state-space structure of the attractors
and basin boundaries in the noise-free self-sustained system. We
will show that the features of birhythmicity  in this modified van
der Pol oscillator strongly depend on $\alpha$ and $\beta$.

\section{Dynamical attractors and birhythmicity properties}
\noindent

In this Section we summarize the dynamical attractors of the
modified  van der Pol model (\ref{eq1}) without Gaussian noise.
The periodic solutions of Eq.~(\ref{eq1}) can be approximated by
\begin{eqnarray}
\label{eq9} x(t)=A\cos \Omega t.
 \end{eqnarray}
We recall that approximated analytic estimates of the amplitude
$A$ and the frequency  $\Omega $ have been derived in Ref.
\cite{enjieu-chabi-yamapi-woafo}, and it has been found that the
amplitude $A$ is independent of the coefficient $\mu$, that only
enters in the frequency $\Omega$.

 \noindent It appears that,
depending on the values of the parameters $\beta$ and $\alpha$,
the modified van der Pol equation (\ref{eq1}) posses one or three
limit cycles. When three limit cycles are obtained, two of them
are stable and one is unstable, a condition for birhythmicity; the
unstable limit cycle represents the separatrix between the basins
of attraction of the two stable limit cycles. We show in Fig.1 the
 bifurcation lines that contour the
region of existence of birhythmicity in the two parameter phase
space ($\beta$-$\alpha$)
\cite{enjieu-chabi-yamapi-woafo,enjieu-yamapi-chabi}.
 The bifurcation line on the left denotes the passage from a single
 limit cycle to three limit cycles, while the right line denotes the
  reverse passage from three limit cycles to a single solution. At the conjunction,
  a codimension-two bifurcation, or cusp\cite{tutorial}, appears . The first bifurcation
  encountered increasing $\alpha$ corresponds to the saddle-node bifurcation of the outer,
  or larger amplitude cycle, while the second bifurcation occurs in correspondence of a saddle-node
  bifurcation of the inner, or smaller amplitude, cycle. The two frequencies associated to the
  limit cycles are very similar close to the lowest $\alpha$ bifurcation and clearly distinct at
  the highest $\alpha$ bifurcation line, as will be discuss later in detail.

Table $1$ provides, for some selected sets $S_i$ of the parameters
in the domain of existence of three limit-cycles on which we will
focus our attention, the comparison between amplitudes and
frequencies derived from the analytical estimate of
Ref.\cite{enjieu-chabi-yamapi-woafo} and from numerical
simulations of Eq.~(\ref{eq1}). From the Table it is clear that
birhythmicity is indeed present -- the two stable attractors are
characterized by different frequencies. However, the two
frequencies are very similar, and in practice it might prove very
difficult to resolve the difference. To illustrate the dynamics of
the self-sustained oscillations, we report in Fig.2 the limit
cycles and in Fig.3 and 4 the time dependent oscillations. In Fig.
3, the two frequencies are very similar, while in Fig. 4 we report
the case of two clearly distinct frequencies. It is clear that for
the slow oscillations (the solid line in Fig. 4, the behavior is
not well approximated by the sinusoidal approximation (\ref{eq9}).
It can also be noticed that the amplitude is still captured by the
theory, while the agreement between the predicted and the observed
frequency becomes poor at low frequencies. In fact for Fig. 4(i),
$\alpha=0.12$, $\beta=0.0014$, the theoretical analysis
\cite{enjieu-chabi-yamapi-woafo} predicts $A_1=2.49$ and
$A_3=10.89$, with frequencies $\Omega_1=0.999$ and
$\Omega_3=0.532$, respectively, in good agreement with the
numerical data $\Omega_1=1.00$ and $\Omega_3=0.516$. For the case
of Fig. 4(ii), $\alpha=0.13$, $\beta=0.001$, the theoretical
analysis \cite{enjieu-chabi-yamapi-woafo} gives $A_1=2.828$ and
$A_3=13.84$, with frequencies $\Omega_1=0.998$ and
$\Omega_3=0.521$, while the numerical data read $\Omega_1=1.00$
and $\Omega_3=0.195$. It is evident that the observed frequency of
the large cycle, $0.195$, is much less than the predicted value
$0.521$.

In order to understand the effect of the parameters $\alpha$ and
$\beta$ on the dynamical states, we have simulated Eq.~(\ref{eq1})
to numerically derive the frequencies $\Omega_i$; the results are
shown in Table $2$. For $\alpha$ and $\beta$ in the white area of
Fig.1, there exists only a single limit-cycle solution. In the
gray area of Fig. 1 there are multi-limit-cycle solutions with
$\Omega_1 \neq \Omega_3$. Fig. 5 shows the dependence of the
frequencies $\Omega_i$ versus the coefficient $\beta$ when the
parameter $\alpha$ is fixed. In this parameter region for each
value of $\alpha$, the two limit-cycle frequencies are different
at low $\beta$ values (see Fig. 4), but converge to the same
frequency when $\beta$ increases (see Fig. 3). This reveals that
the
 saddle-node bifurcation at the upper boundary of the multi-limit-cycles
  area in Fig. 1 occurs when the two frequencies are very similar. Thus we
conclude that birhythmicity smoothly disappears increasing $\beta$
because the two frequencies become undistinguishable, while the
attractors are clearly distinct at the saddle-node bifurcation.

Fig. 6 shows the dependence of $\Omega_i$ versus $\alpha$ for
different values of $\beta$. As $\alpha$ increases, we move from
the boundaries of the multi-limit-cycle area where $\Omega_1 =
\Omega_3$ to enter the region of the map in which the two
limit-cycle frequencies are different ({\it i.e.} $\Omega_1 \neq
\Omega_3$).

 So we conclude that the saddle-node bifurcation at the right hand side
  refers not only to the appearance of a new limit cycle, but also to a
  cycle with a definitely different frequency, and therefore in this
  region birhythmicity is more easily observed. In contrast, it
is evident that it will be extremely difficult to detect
birhythmicity for low $\alpha$.

\section{Numerical estimate of escape rates and global stability analysis}
\subsection{Escape times from the periodic attractors}
\noindent

At non zero noise intensity ($D\neq 0$), the random force causes
the system to occasionally jump from one limit cycle to the other.
The system initialized on a given limit-cycle attractor (with
amplitude $A_1$ or $A_3$) is forced by the random fluctuations of
the $\Gamma$ term in Eq.~(\ref{eq2}) to leave the attractor and to
wander about in the neighboring state space. Escape occurs when
this random motion drives the system across the boundary of the
basin of attraction ({\it i.e.} across the unstable limit-cycle
with amplitude $A_2$). The mean time $\tau$ required for escape
from a basin of attraction is a useful measure of the attractor's
global stability. This escape time is analogous to the escape time
of a system trapped in a minimum of the effective potential, and
the escape implies that the random force drives the system to the
other minimum of the effective potential. The activation energies
shown in Fig. 7 sketch the escape process to be considered in the
following subsection. In fact there are two metastable states:

\begin{enumerate}
\item The  system is trapped at the effective potential minimum in
the basin of attraction of the limit-cycle amplitude $A_1$. Then,
escape to the basin of attraction with limit-cycle amplitude $A_3$
occurs when the system under Gaussian white noise crosses the
unstable limit-cycle amplitude $A_2$ ({\it i.e.} $|x|>A_2$). This
can be numerically computed by choosing the initial conditions
close to the origin. Thus, the corresponding effective energy
barrier to escape from the basin of attraction with limit-cycle
amplitude $A_1$ to the one with amplitude $A_3$ is called $\Delta
U_1$.

\item In the reverse situation, the system is trapped at the
effective potential minimum in the basin of attraction of the
limit-cycle amplitude $A_3$. The initial conditions are chosen
outside the basin of attraction of the limit cycle $A_1$ and far
of the unstable limit-cycle $A_2$. We will denote with $\Delta
U_3$, the effective energy barrier to escape from the basin of
attraction with limit-cycle amplitude $A_3$ across the unstable
limit cycle with amplitude $A_2$ ({\it i.e. $|x|<A_2$}) towards
the limit-cycle with amplitude $A_1$.
\end{enumerate}

Fig. 7 sketches our notation and the most relevant cases:

\begin{itemize}
\item   Case {\bf (i)}: Fig. 7(i) corresponds to the case where
$\Delta U_1$ is larger than $\Delta U_3$. We shall see that
$\Delta U_1$ can became very large (depending on the coefficients
$\alpha$ and  $\beta$); in such conditions the attractor of the
limit-cycle amplitude $A_1$ is much more stable than the
limit-cycle amplitude $A_3$. Thus, the system is more likely to
stay on the limit-cycle attractor $A_1$.

 \item   Case {\bf (ii)}: Fig.
 7(ii) depicts the symmetric case $\Delta U_1 \simeq \Delta
 U_3$. Both attractors are equivalent and we are in
 a  symmetric bistable double well. The system has
 approximately the same probability to stay in one or the other basins.
 \item  Case {\bf(iii)}:
 Fig. 7(iii) shows the case where the energy barrier $\Delta
U_1$ is less than $\Delta U_3$. Here, is the reverse situation of
the case (i), and the first attractor is less stable. the system
is more likely to stay on the limit-cycle attractor $A_3$.
\end{itemize}
Thus, while in principle bistability occurs for all values of the
parameters $\alpha$ and $\beta$ in the gray area of Fig. 1, noise
driven bistability  is more likely to be observed in a narrower
region of the parameter space, see case (ii).


\subsection{Numerical estimate of the escape rates and effective energy barriers}
\noindent

Although there exists a method for the calculation of activation
energies in non-equilibrium systems that do not admit a bona fide
potential using the principle of minimum available noise energy
\cite{dykman78,dykman01,kautz,freidlin,graham81,graham85}, we
adopt here the indirect approach of computing the escape time and
then we infer on the values of the activation energies. The mean
escape time $\tau$ is computed as the average over a series of
trials of the time $\tau_i$ required for the system to move from
one attractor to the other attractor under the influence of noise.
For each trial, integration is begun at $t=0$ with the system
initialized on the attractor and proceeds  by numerically solving
the system equations with a finite difference integration method
of step size $\Delta t$ (see Eq.~(\ref{euler})). The fact that the
random motion of the system is due to a Gaussian white noise
ensures that escape will occurs with probability $1$ within a
finite time \cite{kautz}. Thus, the main question is how long the
system stays in the same basin of attraction. We expect that the
escape time is given by the inverse  Kramer escape rate, or from
the  Arrhenius factor \cite{kramer}:
\begin{equation}
\label{eqtau} \tau \simeq \exp(\Delta U_i/D),
\end{equation}
where $\Delta U_i$ (i=1,3) is the difference between maximum and
minimum values of an effective potential.


We remark that a function plays the role of a thermodynamic
potential for fluctuating dissipative systems that do not possess
a bona fide potential \cite{graham85} if it correctly describes
the asymptotic response to noise. In a sense, one reverses the
Kramer logic: it is called effective potential a function $U$ that
gives the slope of the logarithm of the escape time vs the inverse
of the noise intensity for low noise strength (see
Eq.(\ref{eqtau}) ): $U \propto log(\tau/D)$ (for $D \rightarrow
0$). In this framework, one could regard the potential $U$ as a
way to summarize the behavior of the escape times. In other words
it is completely equivalent either to say that the escape times
are exponentially distributed vs the inverse of the noise (for low
noise) with slope $U$ or that the effective potential reads $U$.

The relevant attractors and basins of attraction are those shown
in Figs.2.
 The data show that
the mean escape times $\tau$ obtained from simulations for both
limit-cycle state $A_1$ and $A_3$ state increase exponentially
with the inverse noise intensity. With the parameter sets $S_i$,
we find that the variation of the average escape time (on a
logarithm scale) as function of the inverse noise intensity $1/D$
strongly depends to the nonlinear coefficients $\alpha$ and
$\beta$.
 For example, the
sets $S_1$, $S_4$, and $S_6$ correspond to case (i) in which the
attractor of the limit-cycle $A_3$ is less stable than the
attractor $A_1$. The symmetric bistable situation, case (ii) is
observed with the set $S_5$. The last case (iii) is found for the
sets $S_2$ and $S_5$. It is important to note that the case (ii)
only occurs in a very narrow range, $0.08 < \alpha < 0.09$ and
$0.0012 < \beta < 0.0014 $ \cite{dykman78,dykman01}. Outside this
narrow area the properties of the two attractors are very
different.

 {\scriptsize
\begin{eqnarray*}
\begin{tabular}{|l|c|c|c|c|c|l|}
\hline
& $\alpha={\bf 0.07}$ & $\alpha={\bf 0.08}$ & $\alpha={\bf 0.09}$ & $\alpha={\bf 0.1}$ &$\alpha={\bf 0.12}$&$\alpha={\bf 0.13}$\\
\hline
$\beta={\bf 0.004} $         &    &     &        &        && $\Delta U_1$={\bf 0.074} \\

                        &     &   &   &       & &$\Delta U_3$= {\bf 0.0072}\\
\hline
$\beta={\bf 0.003} $         &    &     &        &        &$\Delta U_1$= {\bf 0.095}& $\Delta U_1$={\bf 0.028} \\

                        &     &   &   &       &$\Delta U_3$={\bf 1.656} &$\Delta U_3$= {\bf 0.0075}\\
\hline
$\beta={\bf 0.0025} $         &    &     &        &        &$\Delta U_1$= {\bf 0.054}& $\Delta U_1$={\bf 0.015} \\

                        &     &   &   &       &$\Delta U_3$={\bf 2.7} &$\Delta U_3$= {\bf 6.75}\\
\hline
$\beta={\bf 0.002}$           &     &    &       & $\Delta U_1$= {\bf 0.25}    &$\Delta U_1$={\bf 0.035 } & $\Delta U_1$= {\bf 0.0097}   \\

                        &     &     &       & $\Delta U_3$=  {\bf 0.75}   &$\Delta U_3$={\bf 10.5 } &$\Delta U_3$={\bf 28.8} \\
\hline

$\beta={\bf 0.0016}$          &     &    & $\Delta U_1$={\bf 0.45}  & $\Delta U_1$= {\bf 0.183}  &$\Delta U_1$= {\bf 0.026}& $\Delta U_1$={\bf 0.0035}   \\

                        &     &    & $\Delta U_3$= {\bf 0.93}  &$\Delta U_3$=  {\bf 7.78}  &$\Delta U_3$= {\bf 68.2}&   $\Delta U_3$={\bf 224} \\
\hline
$\beta={\bf 0.0014}$          &     & $\Delta U_1$= {\bf 0.98}   & $\Delta U_1$= {\bf 0.34}  & $\Delta U_1$ = {\bf 0.16}  &$\Delta U_1$= {\bf 0.021} & $\Delta U_1$= {\bf 0.0017}    \\

                        &     & $\Delta U_3$= {\bf 0.014}   & $\Delta U_3$= {\bf 3.78}  & $\Delta U_3$= {\bf 16.14} &$\Delta U_3$= {\bf 152.3}& $\Delta U_3$= {\bf 233.5}\\
\hline
$\beta={\bf 0.0012}$          &    & $\Delta U_1$= {\bf 0.62}    & $\Delta U_1$= {\bf 0.291} & $\Delta U_1$= {\bf 0.13}  &$\Delta U_1$= {\bf 0.104}&$\Delta U_1$={\bf 0.0015}    \\

                        &    & $\Delta U_3$={\bf 2.15}     & $\Delta U_3$= {\bf 11.6}  & $\Delta U_3$= {\bf 17.5} &$\Delta U_3$={\bf 308} &$\Delta U_3$= {\bf 791}\\
\hline

$\beta={\bf 0.0011}$          &    &  $\Delta U_1$= {\bf 0.65}  & $\Delta U_1$={\bf 0.28}   & $\Delta U_1$= {\bf 0.123} & $\Delta U_1$= {\bf 0.015 } &$\Delta U_1$= {\bf 0.003} \\

                        &   & $\Delta U_3$={\bf 4.35}    & $\Delta U_3$={\bf 27.5}    & $\Delta U_3$= {\bf 104.9} & $\Delta U_3$= {\bf 564} &$\Delta U_3 >$ {\bf 1000 }\\
\hline

$\beta={\bf 0.001}$                & $\Delta U_1$={\bf 1.3} & $\Delta U_1$=  {\bf 0.52}  & $\Delta U_1$= {\bf 0.25 }  & $\Delta U_1$={\bf 0.11} &$\Delta U_1$= {\bf 0.014} &   $\Delta U_1$={\bf 0.0001}    \\

                             & $\Delta U_3$={\bf 0.53}      & $\Delta U_3$=   {\bf 10.7}      & $\Delta U_3$=  {\bf 16.05}     & $\Delta U_3$=  {\bf 105.6} &$\Delta U_3 >${\bf 1000}&  $\Delta U_3 >$ {\bf 1000 }\\
\hline

\end{tabular}
\end{eqnarray*}}
{\it Table 2: \small Dependence of the energy barriers $\Delta U_i$ in the parameters plane $(\alpha,\beta)$, with $\mu=0.1$.}\\

Fitting a straight line through the data points in the linear part
of Eq.(\ref{eqtau}) and measuring its slope we obtain an estimate
of $\Delta U_1$ and $\Delta U_3$, the effective activation
energies for the escape from the limit-cycle attractor $A_1$ and
$A_3$, respectively. Since the effective activation energy is
defined by the low-noise intensity asymptote, the accuracy of
numerical simulation estimates can be affected if high-noise
intensity points ({\it i.e.}, points where the relation is not
linear) are included in the fitting procedure. For this reason,
data points for which the resulting Arrhenius factor bends have
been excluded from the fitting procedure (we employ a $\chi^2$
test to check for linearity). Fig. 8 shows the variation of the
effective energy barriers versus the coefficient $\mu$ with the
set of parameters $S_i$. The effective energy barriers increase
when $\mu$ increases, and the behaviors strongly depend upon the
set of the parameters $S_i$. The scenarios mentioned in subsection
IV-B  can be found in the behaviors of $\Delta U_{1,3}$ ({\it
i.e.} cases (i), (ii), (iii)). The case (i) appears in Fig. 8 for
the sets $S_1, S_4, S_6$, in which the energy barrier $\Delta U_1$
quickly increases. Here, one concludes that the limit-cycle
attractor $A_1$ of the modified van der Pol oscillator is much
more stable than the attractor $A_3$ (respect to  Gaussian white
noise). The system will likely stay for a long time in the
effective potential well of the limit-cycle attractor $A_1$, for
the corresponding effective barrier is higher. For instance when
$\mu=0.5$ in $S_1$, we observe $\Delta U_1/\Delta U_3 \simeq 80$.
The set $S_5$ corresponds to the almost symmetric bistable
situation, {\it i.e.} case (ii). Both effective energy barriers
$\Delta U_1$ and $\Delta U_3$ increase when $\mu$ increases and
are comparable: the system remains for approximately the same time
in the two effective potential wells. In the last scenario $S_2$
and $S_3$, {\it i.e.} case (iii), we have a phenomenon opposed to
that of the case (i): the limit-cycle attractor $A_3$ is much more
stable than the attractor $A_1$. The system remains for a much
longer time in the limit-cycle attractor $A_3$ because the energy
barrier is too high, so if the noise level is large enough to
cause a switch from $A_3$ to $A_1$, the same noise will drive back
the system to $A_3$ in a very short time interval with very high
probability.

Let us remark that "short" and "long" might be very different
\cite{dykman78,dykman01,stambaugh06}. To measure the different
properties, we compute the average persistence or residence time
$P_{1,3}$ on the attractor with limit cycle amplitude $A_{1,3}$
as:
\begin{equation}
\label{persistence}
 P_j=\frac{\tau_{j}}{\tau_1+\tau_3},\qquad j=1,3,
\end{equation}
where $\tau_{1,3}$ is the escape time from the first attractor
$A_1$ or third attractor $A_3$, see Eq.~(\ref{eqtau}). For the
parameters $S_1$, for noise intensity around $D=1/20$, we get
$P_3=0.018$, and obviously $P_1=0.982$ {\it i.e.} the system will
spends $1.8 \%$ of the time on the third attractor $A_3$ and
$98.2\%$ on the first attractor $A_1$. Decreasing the noise down
to $D=1/100$, $P_3$ decreases  to $P_3\simeq 3.10^{-9}$. In other
words, for any second spent on the less stable attractor $A_3$ the
system will stay for about $10$ years
on the most stable state $A_1$. Such a dramatic change at low
noise occurs for $\Delta U_1/\Delta U_3\simeq
  50$, from Table $2$ it is clear that ratio between energy barrier can easily be much larger.

To analyze the dynamic structure in the various areas of the chart
drawn on Fig. 1, we present in Table $2$ the effective energy
barriers as a function of the coefficients $\alpha$ and $\beta$
selected in the dotted rectangle of Fig. 1. When $\beta$ is fixed
and $\alpha$ increases, the effective energy barrier $\Delta U_1$
decreases, whereas the energy barrier $\Delta U_2$ considerable
increases. For example, for $\beta=0.0014$, the effective energy
barrier of the limit-cycle attractor $A_1$ decreases from $\Delta
U_1(\alpha=0.08)=0.98$ to the value $\Delta
U_1(\alpha=0.13)=0.0017$, while the
 barrier $\Delta U_3$ increases from $\Delta U_3(\alpha=0.08)=0.014$ to
the value $\Delta U_3(\alpha=0.07)=233.5$. Then, there is a high
probability that the system remains for a longer time in the limit
cycle attractor $A_3$, see Eq.~(\ref{eqtau}). A similar behavior
is reported when $\alpha$ is fixed and that $\beta$ increases. Let
us note about Table $2$ that for low $\beta$ value and high
$\alpha$ values, the case (iii) becomes predominant: $\Delta U_3$
increases and becomes so large that we have not been able to
compute such barrier even with simulations as long as
$t_{max}\simeq 10^{10}$ normalized units. We can only estimate the
barrier to be larger than $1000$.

The behavior of the effective energy barriers can be also
interpreted in the following manner: the right side of the gray
area of existence of bistable regime in Fig. 1, where the two
frequencies are clearly different corresponds to the physical case
where one of the two limit-cycle
 attractors, namely $A_3$, has a very high effective activation energy while
the other, namely $A_1$, vanishes because the effective potential
barrier becomes zero. This process explains the passage from
birhythmicity to a single limit-cycle attractor.

\section{Conclusions}
\noindent

We have considered the characteristics of birhythmicity and the
global stability properties of the attractors in a self-sustained
system. We have found that birhythmicity in a modified van der Pol
oscillator is strongly influenced by the nonlinear coefficients
$\alpha$ and $\beta$: the two frequencies converge or diverge when
the nonlinear coefficients are varied, leading to almost
undistinguishable frequencies for low $\alpha$ and high $\beta$.
Adding a random excitation, we have found that the system crosses
the boundary between the basins of attraction ({\it i.e.} moves
across the unstable limit-cycle with amplitude $A_2$).  The mean
time $\tau$ to escape from one limit-cycle attractor to the other
has been estimated in the low-noise limit, and it is proposed as a
measure of the attractor's global stability. By considering the
variation of the mean escape time $\tau$ versus the inverse noise
intensity $1/D$, the slope of the linear part has enabled us to
summarize the results in the form of an effective activation
energy barrier which is function of the physical system
parameters. We have found, as in other systems that exhibit noise
induced switches between two attractors, that the escape times can
be very different \cite{dykman78,dykman01,stambaugh06}, so it
could be difficult to observe birhythmicity for high $\alpha$ and
low $\beta$. We remark that systems
\cite{dykman78,dykman01,stambaugh06} are periodically driven, and
therefore monorhythmic.

We conclude that although birhythmicity {\it per se } refers just
to the occurrence of two frequencies, actual observation is
subject to much more restrictive conditions. Our purpose is to go
beyond the mere existence of birhythmicity, to show that there are
limitations that restrict the likeliness that birhythmicity
spontaneously occurs. We speculate that there might be other
models that do possess two attractors with different frequencies,
but noise driven  birhythmicity is difficult to  observe because
of the  different stability properties of the attractors. This
might be the reason why birhythmicity has been predicted in many
models, but rarely observed in experiments - actually there is to
our knowledge just one case of clear observation of birhythmic
behavior \cite{bistabexp}. Moreover, the switch from an attractor
to another in Ref. \cite{bistabexp} is due to a change of the
parameters, not to spontaneous transition from a frequency to the
other.
 We
suggest that an analysis similar to that carried out  in this work
is therefore useful to ascertain the birhythmic property in a real
system.


\section*{Acknowledgements}
R. Yamapi undertook this work with the support of the ICTP
(International Centre for Theoretical Physics) Programme for
Training and Research in Italian Laboratories, Trieste, Italy. He
also acknowledges the support of the Laboratorio Regionale
CNR/INFM, Superconducting Materials, Salerno, Italy.


\newpage

\begin{figure}[htb]
\centering
\begin{minipage}{12cm}
\begin{center}
\begin{picture}(150,140)
\put(-20,-40.0){\psfig{file=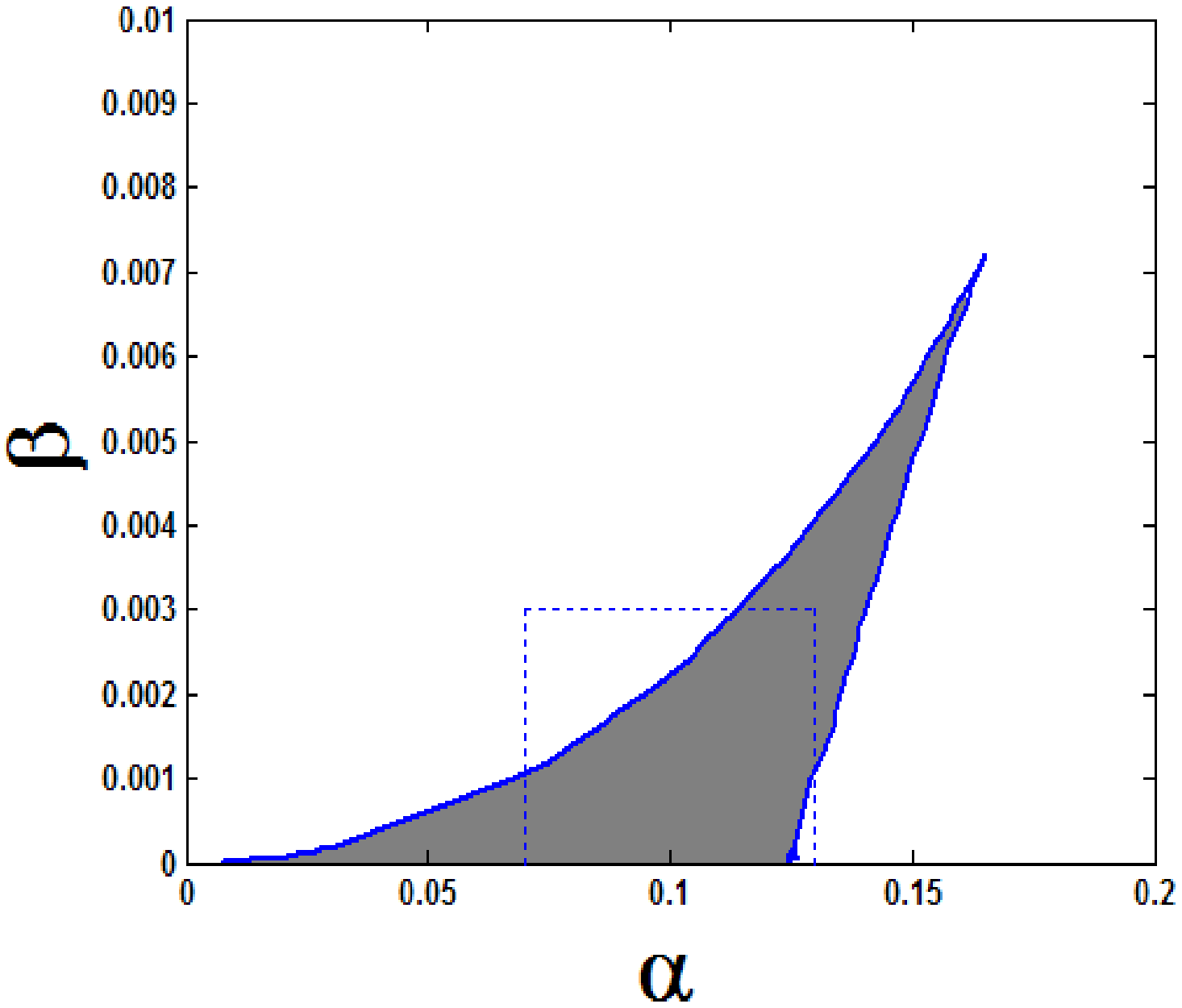,width=16cm,height=15cm,angle=0.0}}
\end{picture}
\caption[] {\footnotesize \it Parameters domain for the existence
of a single limit cycle (white area) and three limit cycles (gray
area) with $\mu=0.1$. The bifurcation line on the left denotes the
saddle-node bifurcation of the outer or large amplitude cycle (see
Fig. 2) while the right hand side contour marks the saddle-node
bifurcation of the inner cycle. The rectangle denotes the
parameter region of Table $2$. } \label{figure1}
\end{center}
\end{minipage}
\end{figure}

\begin{figure}[htb]
\centering
\begin{minipage}{12cm}
\begin{center}
\begin{picture}(250,150)
 \put(0,80){\epsfig{file=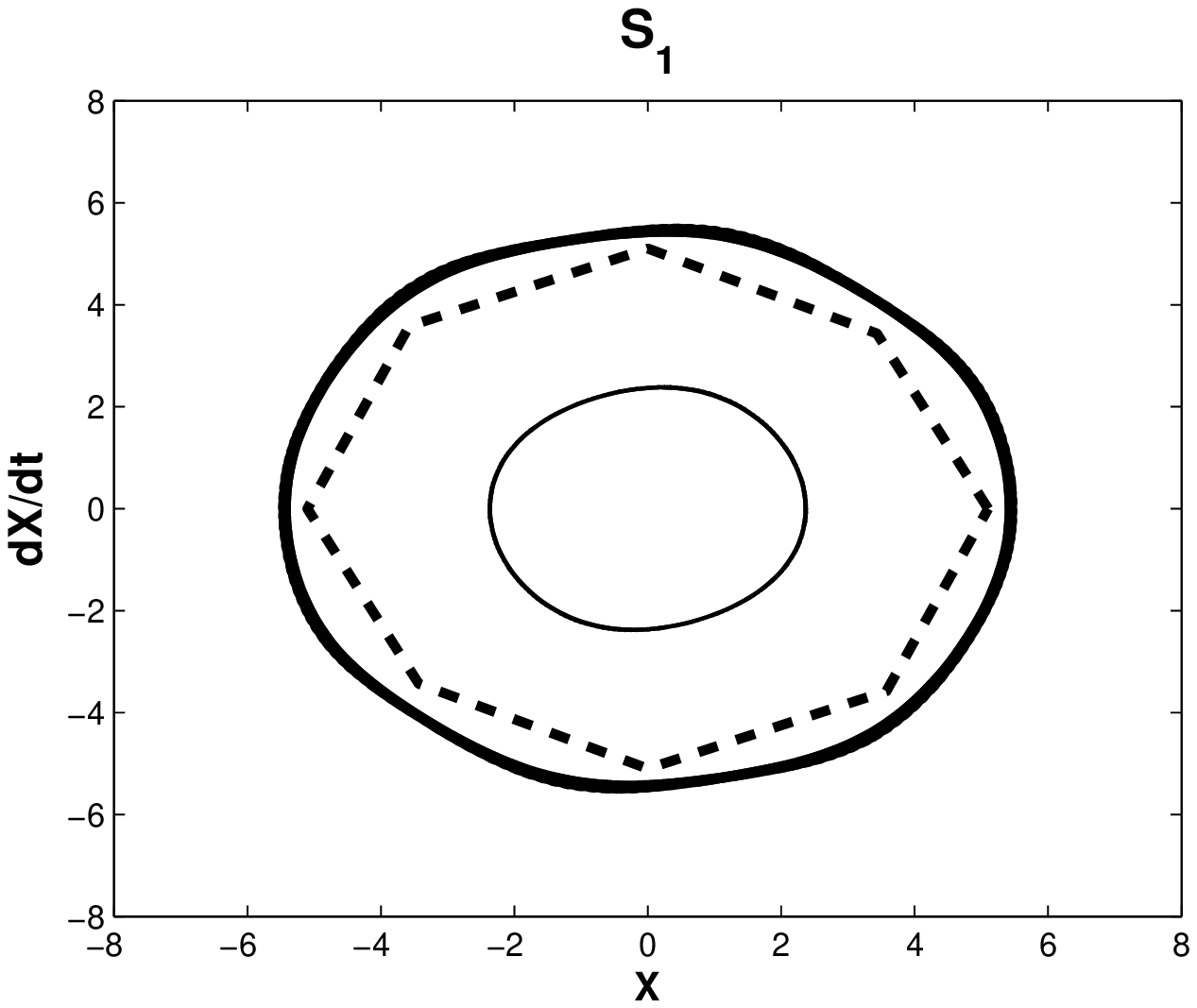,width=6cm,height=4cm,angle=0.0}}
 \put(60,80){\epsfig{file=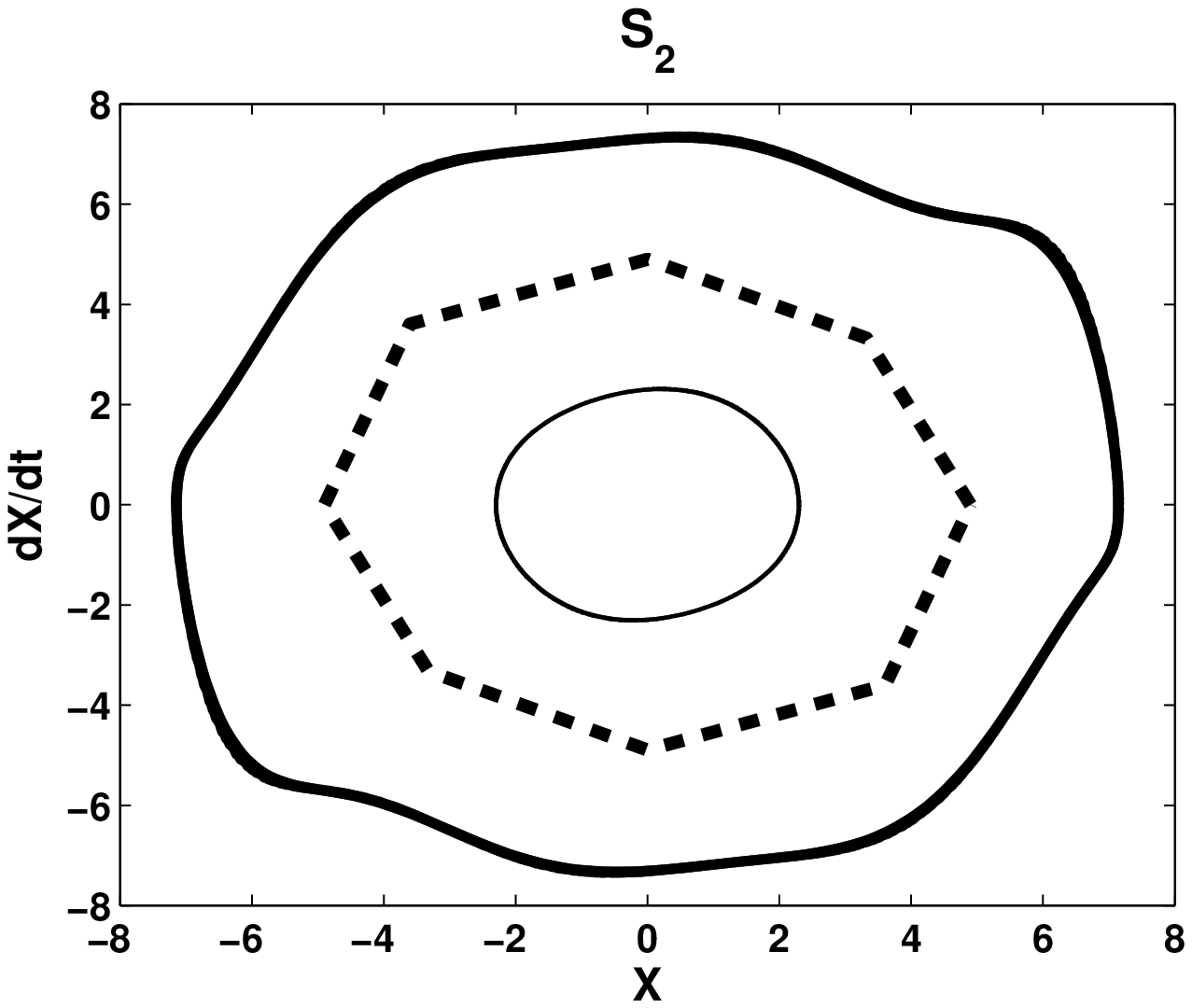,width=6cm,height=4cm,angle=0.0}}
 \put(0,40){\epsfig{file=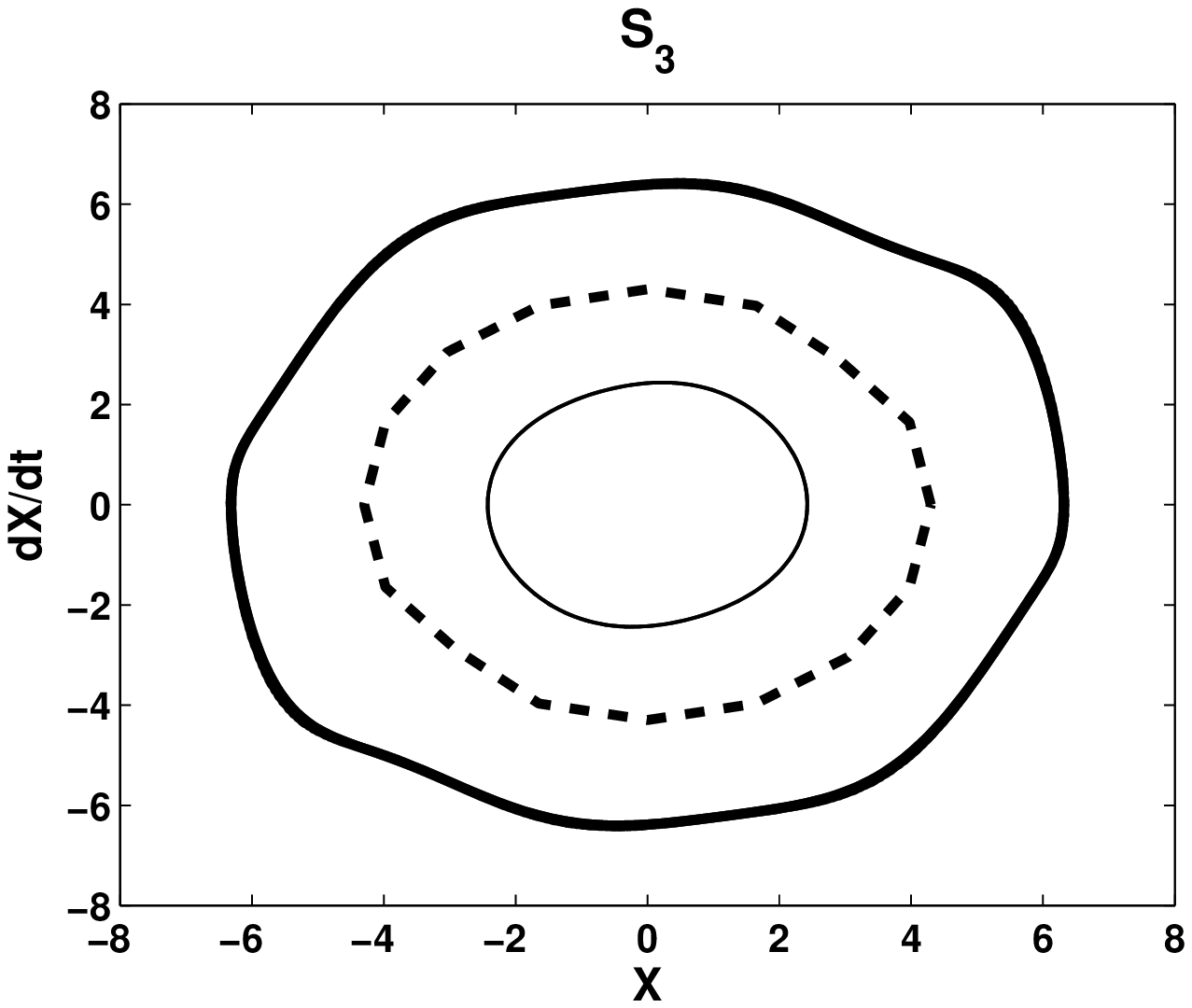,width=6cm,height=4cm,angle=0.0}}
 \put(60,40){\epsfig{file=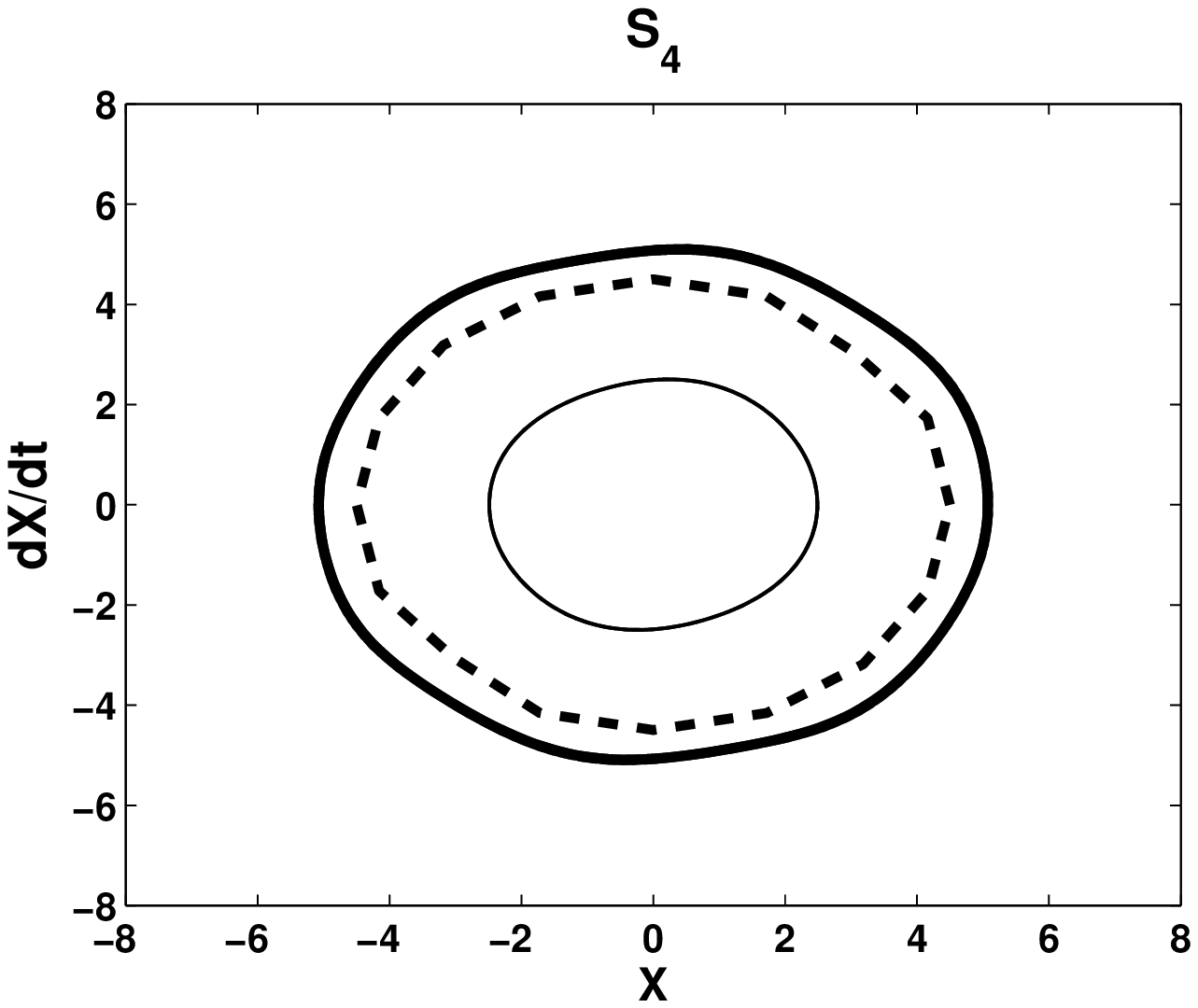,width=6cm,height=4cm,angle=0.0}}
 \put(0,0){\epsfig{file=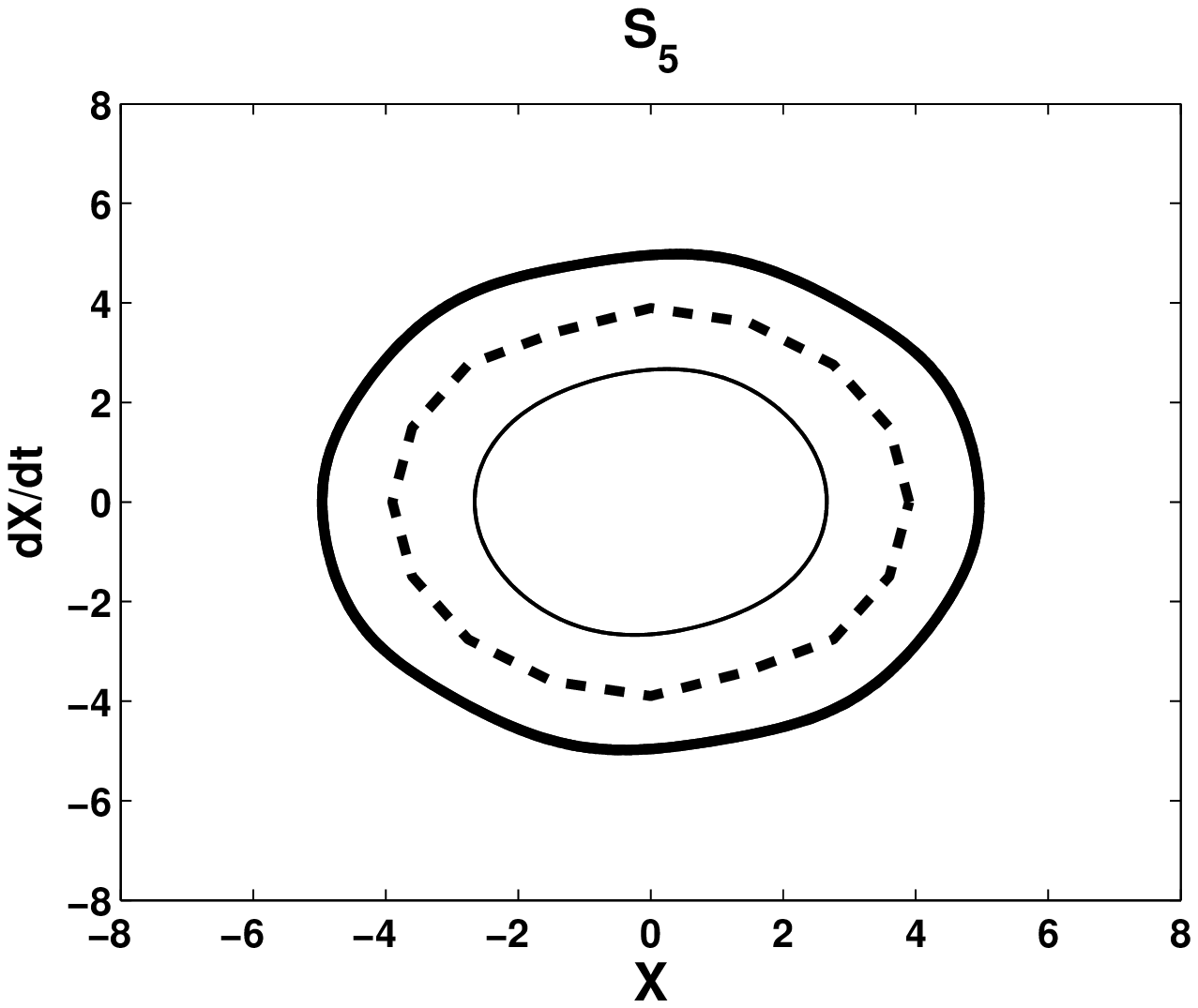,width=6cm,height=4cm,angle=0.0}}
 \put(60,0){\epsfig{file=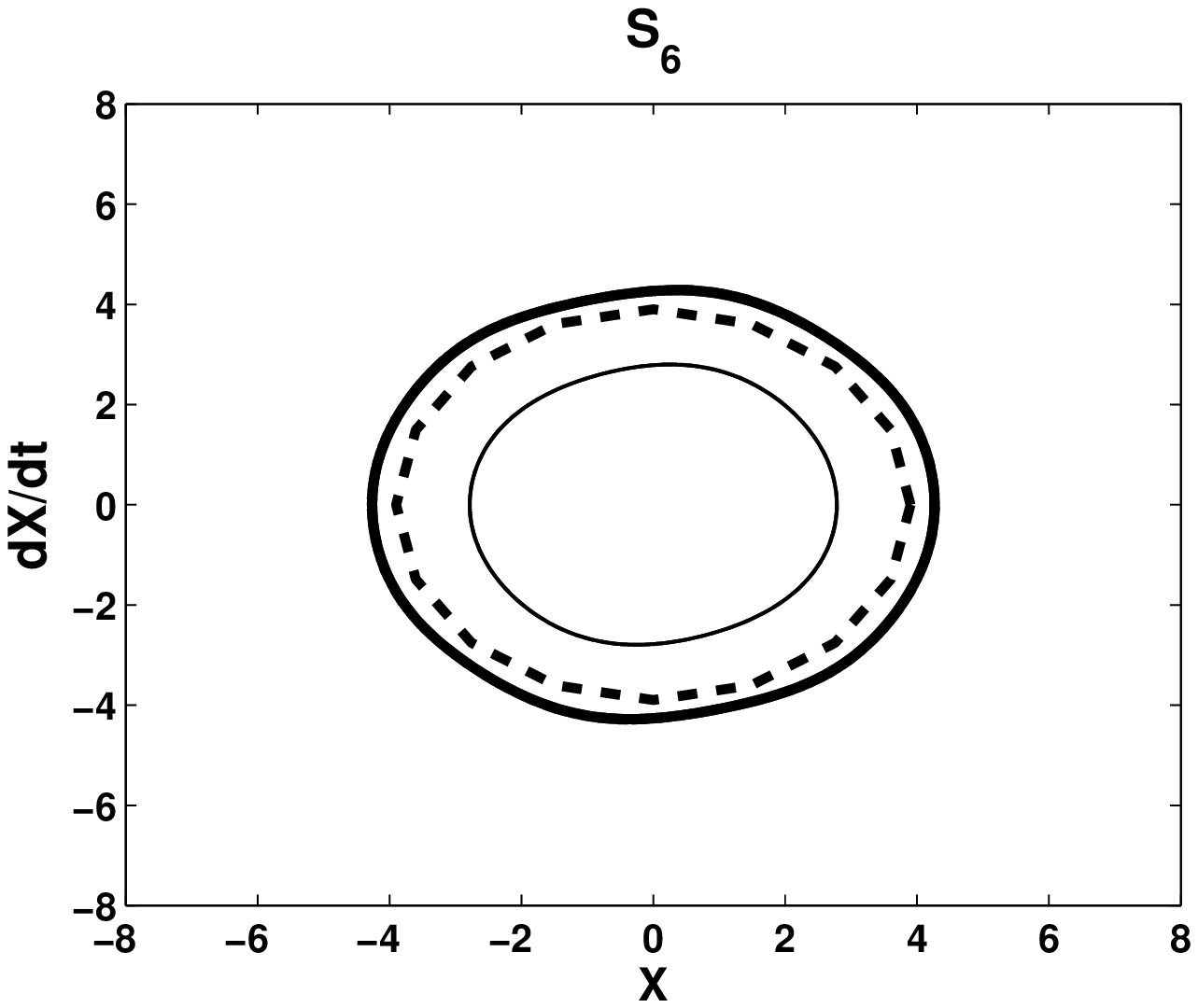,width=6cm,height=4cm,angle=0.0}}
\end{picture}
\caption[] {\footnotesize \it The two stables coexisting limit
cycle attractors obtained by numerical integration of equation
(\ref{eq1})  for $\mu=0.1$ and the sets of parameters $S_i$ (see
Table $1$). The thin line refers to the attractor of smaller
amplitude ($A_1$) and thick line to the larger amplitude ($A_3$).
The dashed line denotes the unstable limit cycle, and separates
the basin of attraction of the inner or smaller amplitude cycle
from the basin of attraction of the outer or larger amplitude
cycle.
 }
\label{figure2}
\end{center}
\end{minipage}
\end{figure}

\begin{figure}[htb]
\centering
\begin{minipage}{12cm}
\begin{center}
\begin{picture}(250,150)
\put(0,80){\epsfig{file=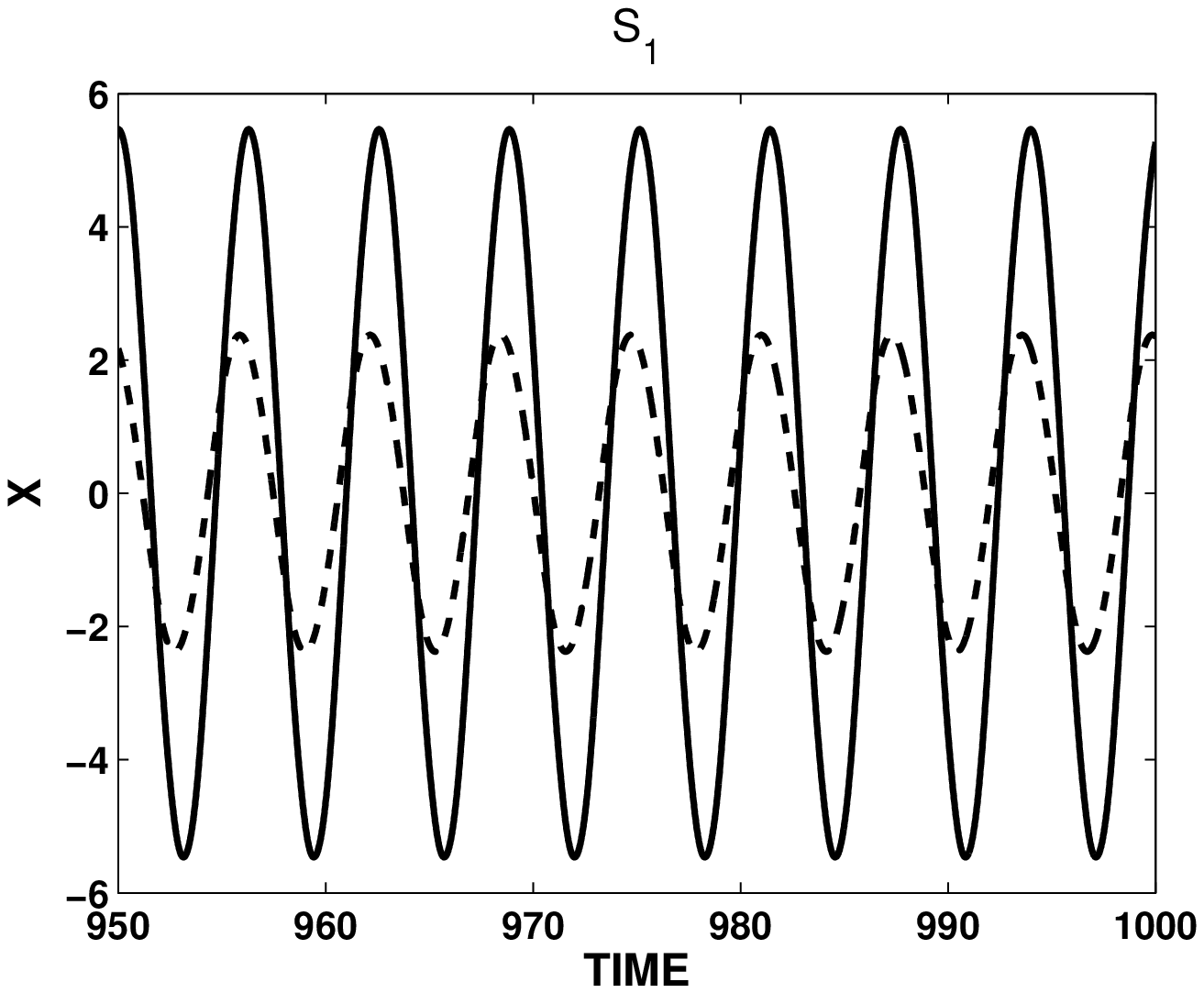,width=6cm,height=4cm,angle=0.0}}
\put(60,80){\epsfig{file=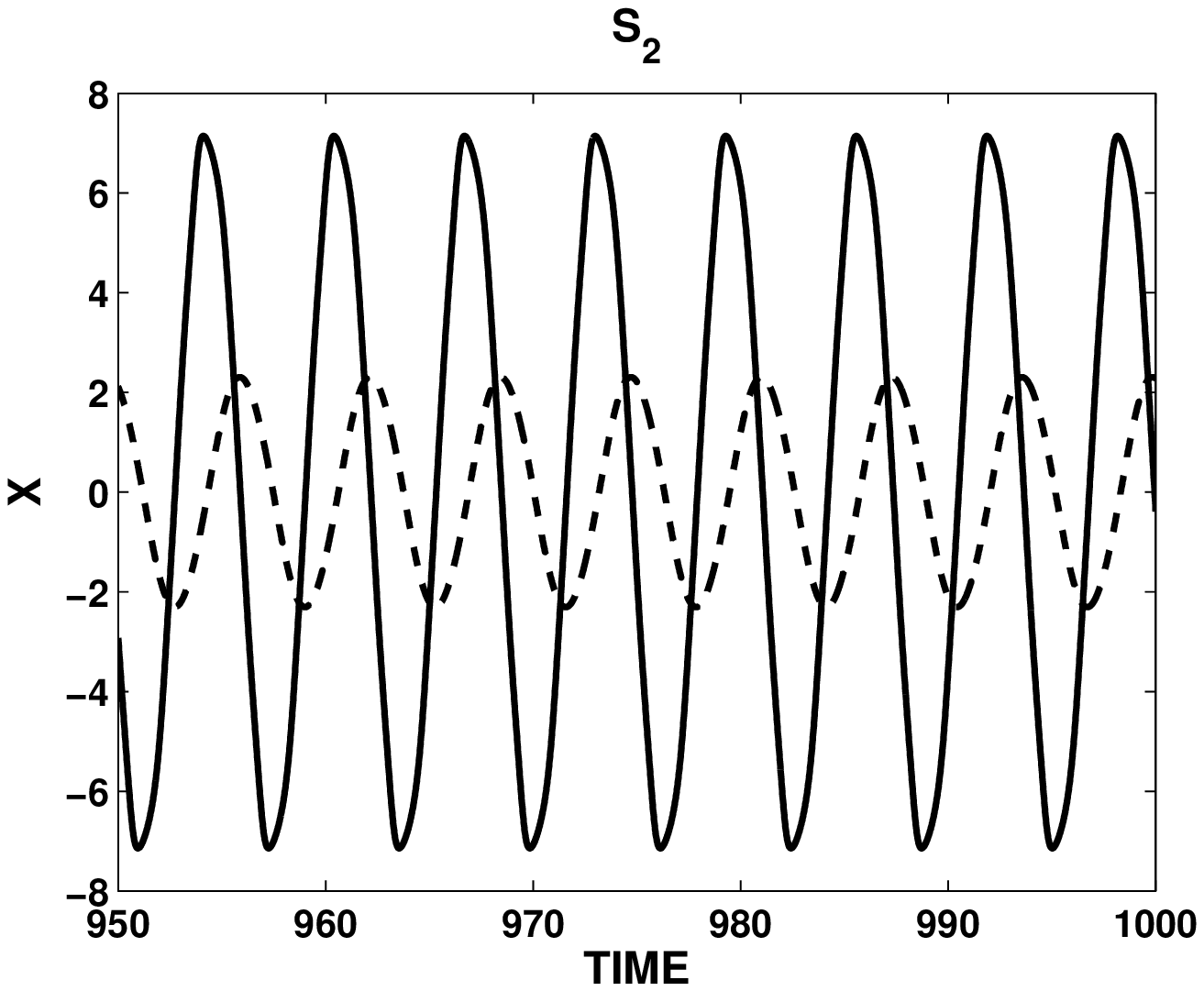,width=6cm,height=4cm,angle=0.0}}
\put(0,40){\epsfig{file=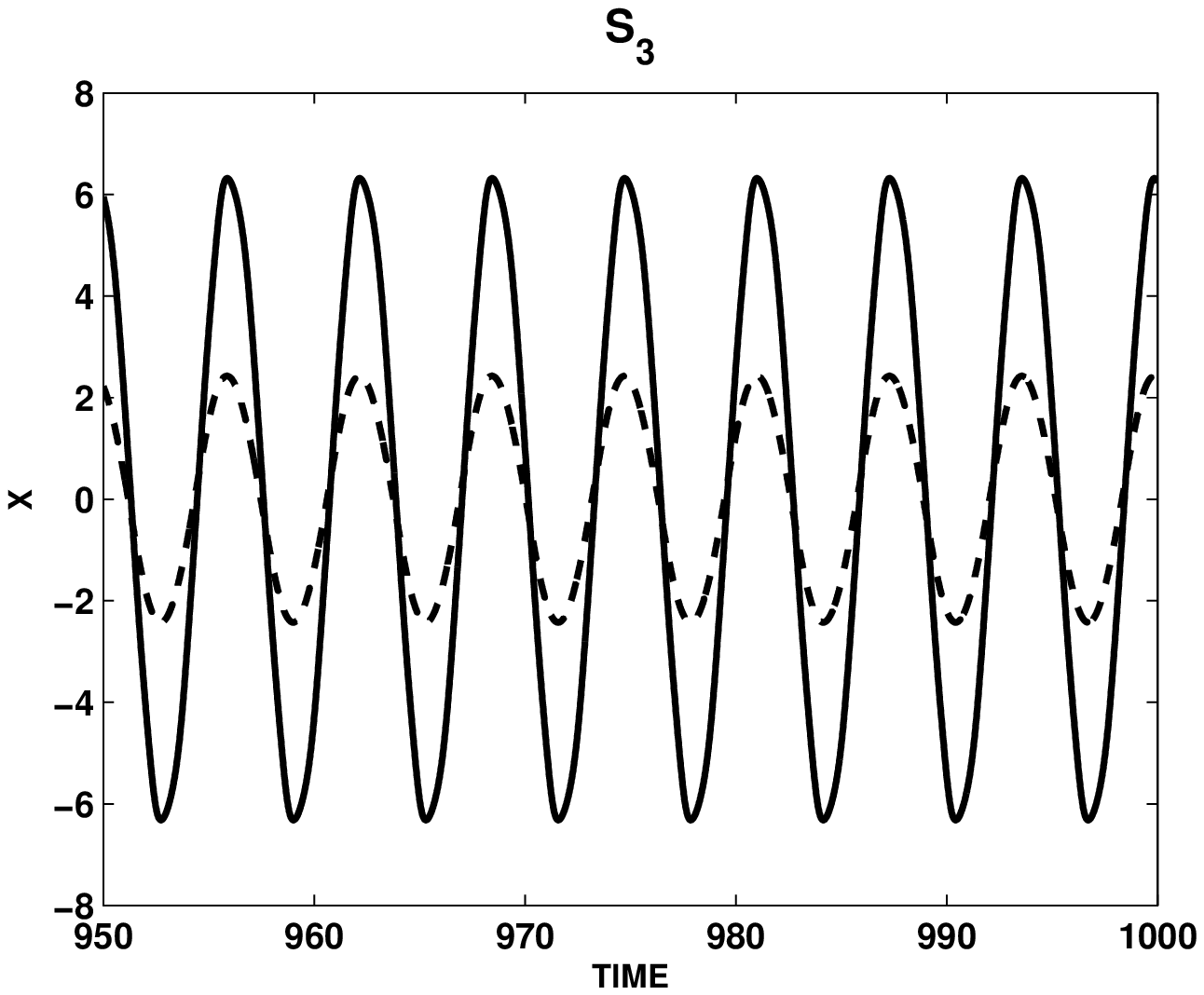,width=6cm,height=4cm,angle=0.0}}
\put(60,40){\epsfig{file=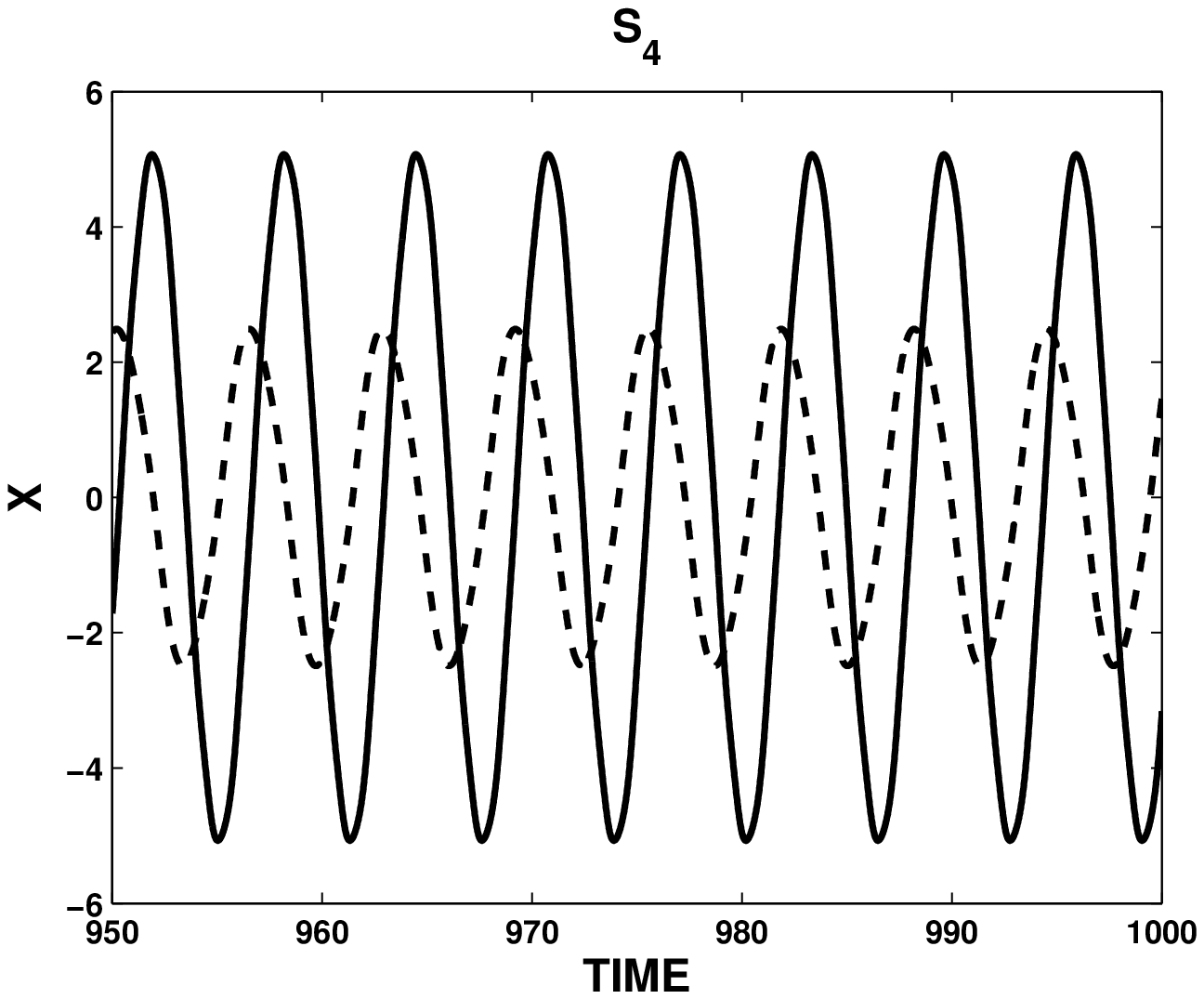,width=6cm,height=4cm,angle=0.0}}
\put(0,0){\epsfig{file=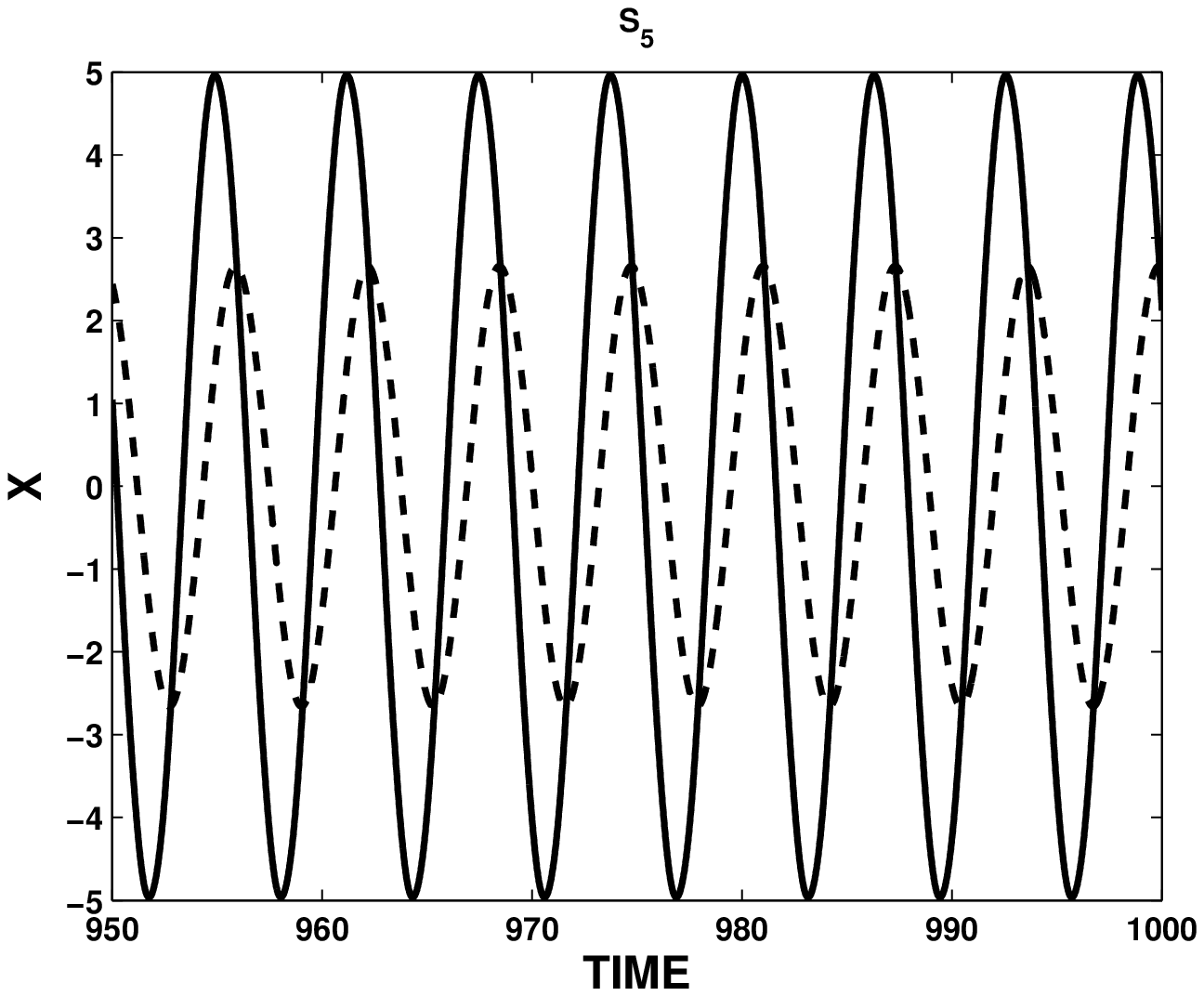,width=6cm,height=4cm,angle=0.0}}
\put(60,0){\epsfig{file=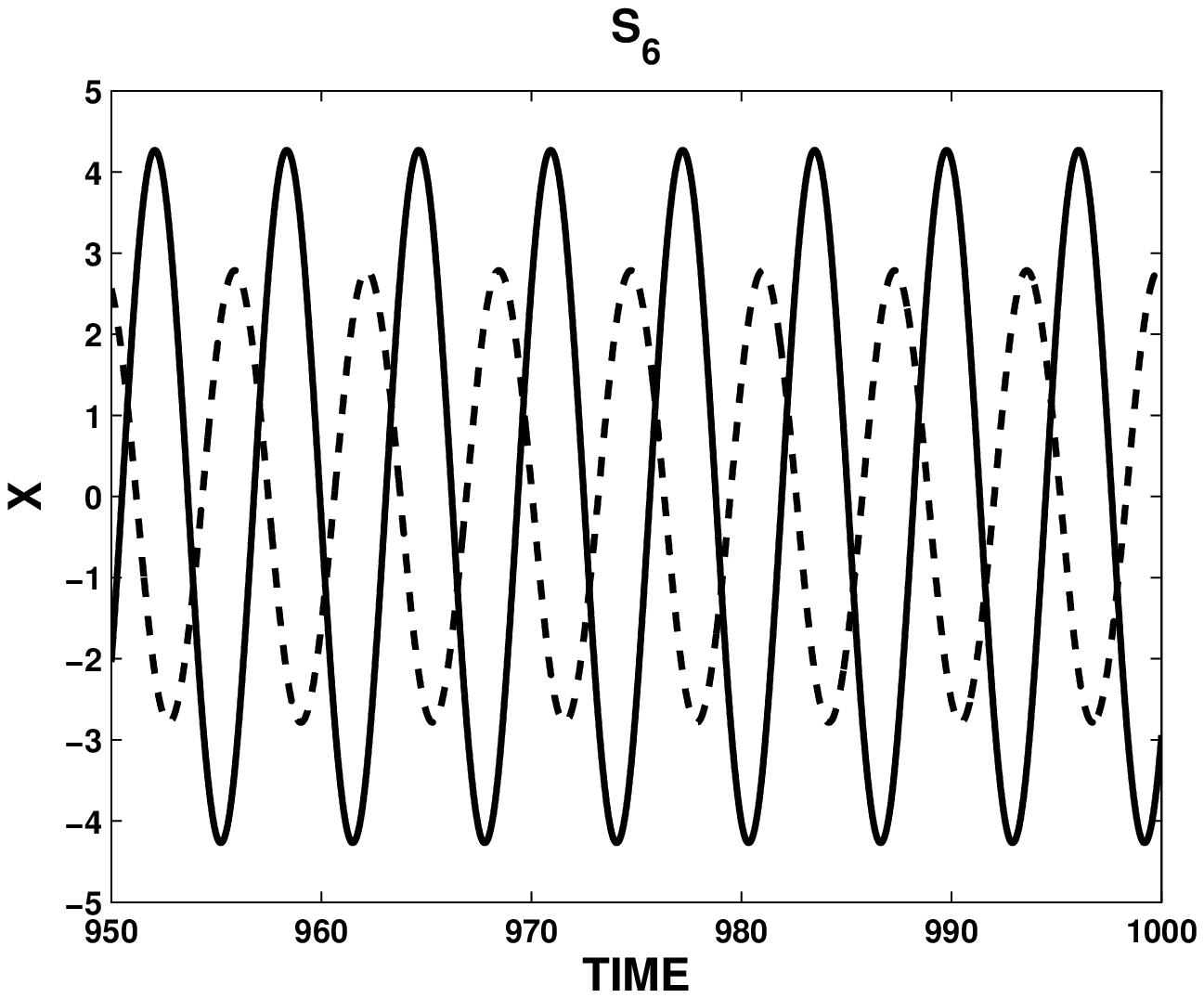,width=6cm,height=4cm,angle=0.0}}
\end{picture}
\caption[] {\footnotesize \it The coexistence between two regimes
of noise-free self-sustained oscillations corresponding to the
larger (solid line, $A_3$, and frequency $\Omega_3$) and smaller
(dotted line, $A_1$, and frequency $\Omega_1$) limit cycles, the
two frequencies are approximately the same, $\Omega_1 \simeq
\Omega_3$. The set of parameters is the same as for the attractors
shown in Fig.\ref{figure2}, see Table $1$.} \label{figure3}
\end{center}
\end{minipage}
\end{figure}

\begin{figure}[htb]
\centering
\begin{minipage}{12cm}
\begin{center}
\begin{picture}(250,140)
\put(10,55){\epsfig{file=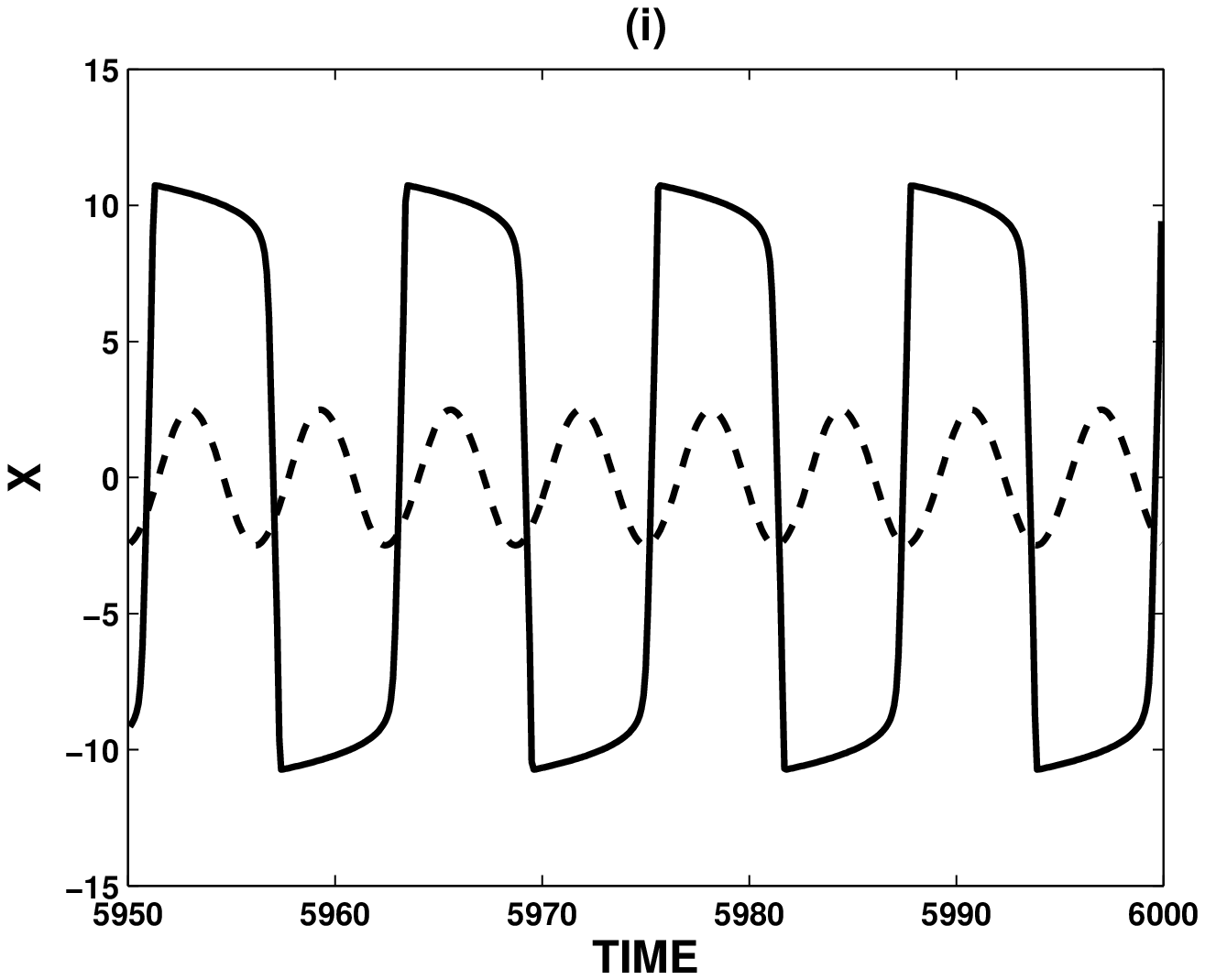,width=10cm,height=5cm,angle=0.0}}
\put(10,0){\epsfig{file=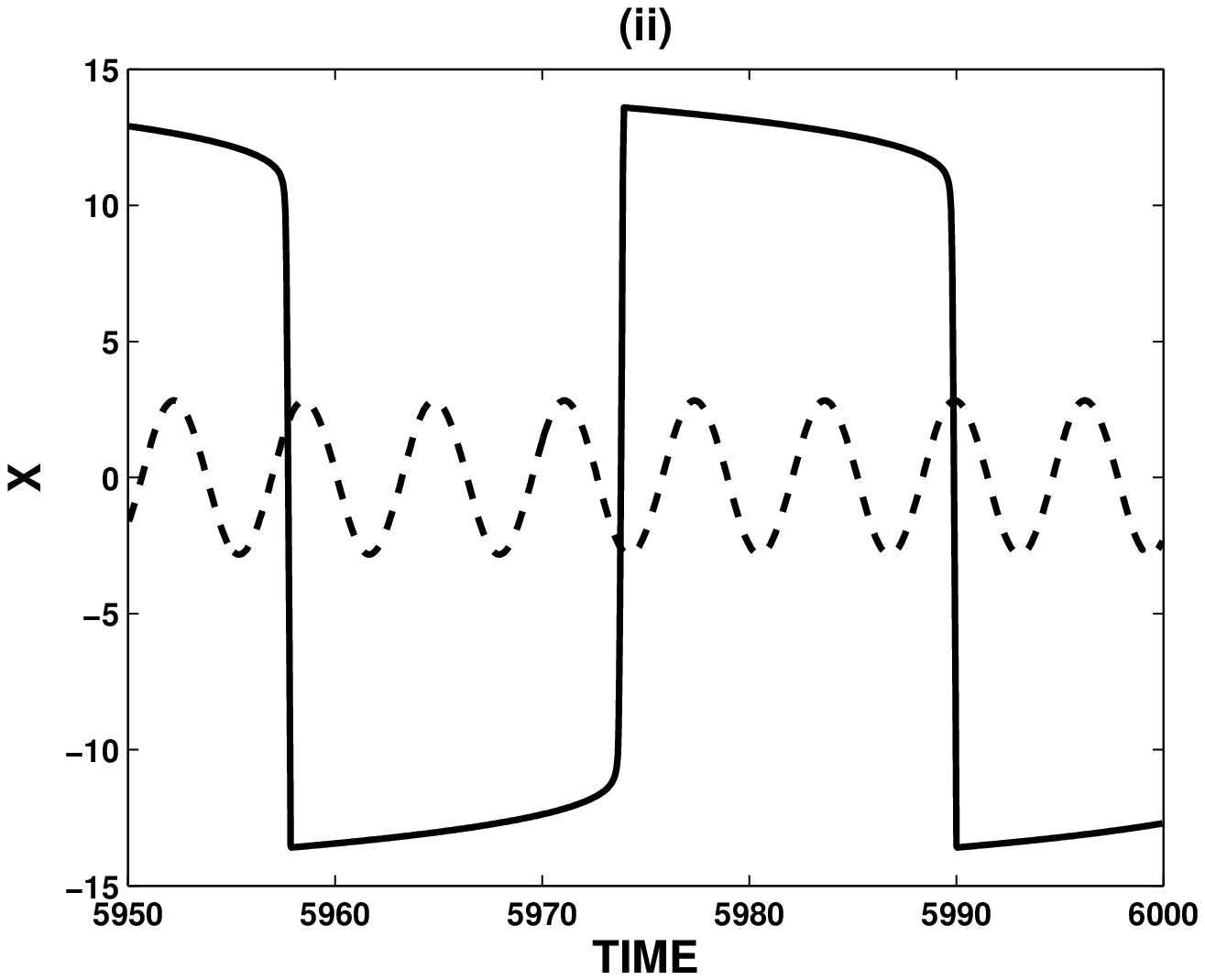,width=10cm,height=5cm,angle=0.0}}
\end{picture}
\caption[] {\footnotesize \it Birhythmicity with oscillations at
two clearly different frequencies, $\Omega_1 =2 \Omega_3$. (i)
$(\alpha;\beta)=(0.12;0.0014)$ and (ii)
$(\alpha;\beta)=(0.13;0.001)$. In both figures we have set $\mu =
0.1$.} \label{figure4}
\end{center}
\end{minipage}
\end{figure}

\begin{figure}[htb]
\centering
\begin{minipage}{12cm}
\begin{center}
\begin{picture}(250,120)
\put(-10,50){\epsfig{file=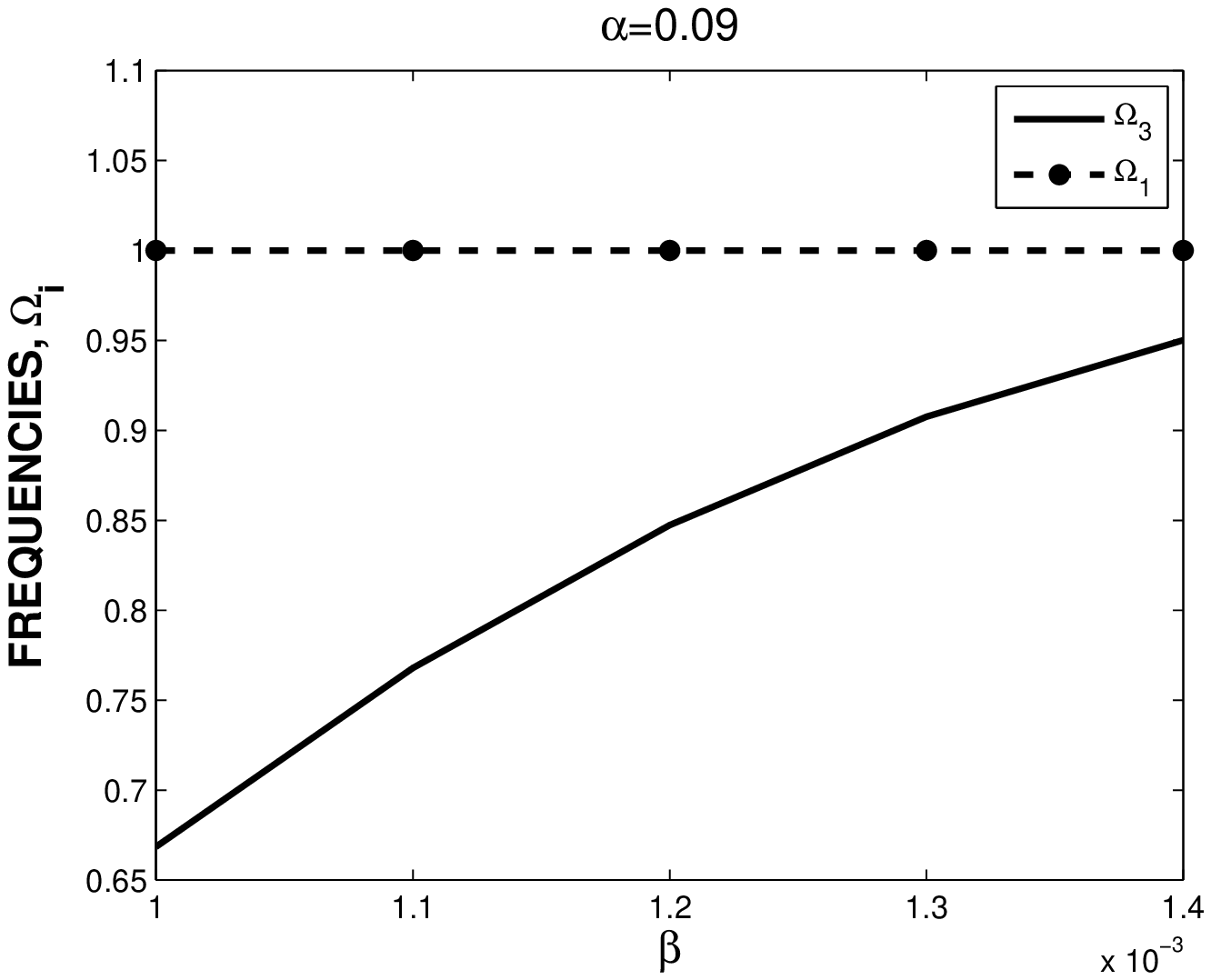,width=7.5cm,height=5cm,angle=0.0}}
\put(70,50){\epsfig{file=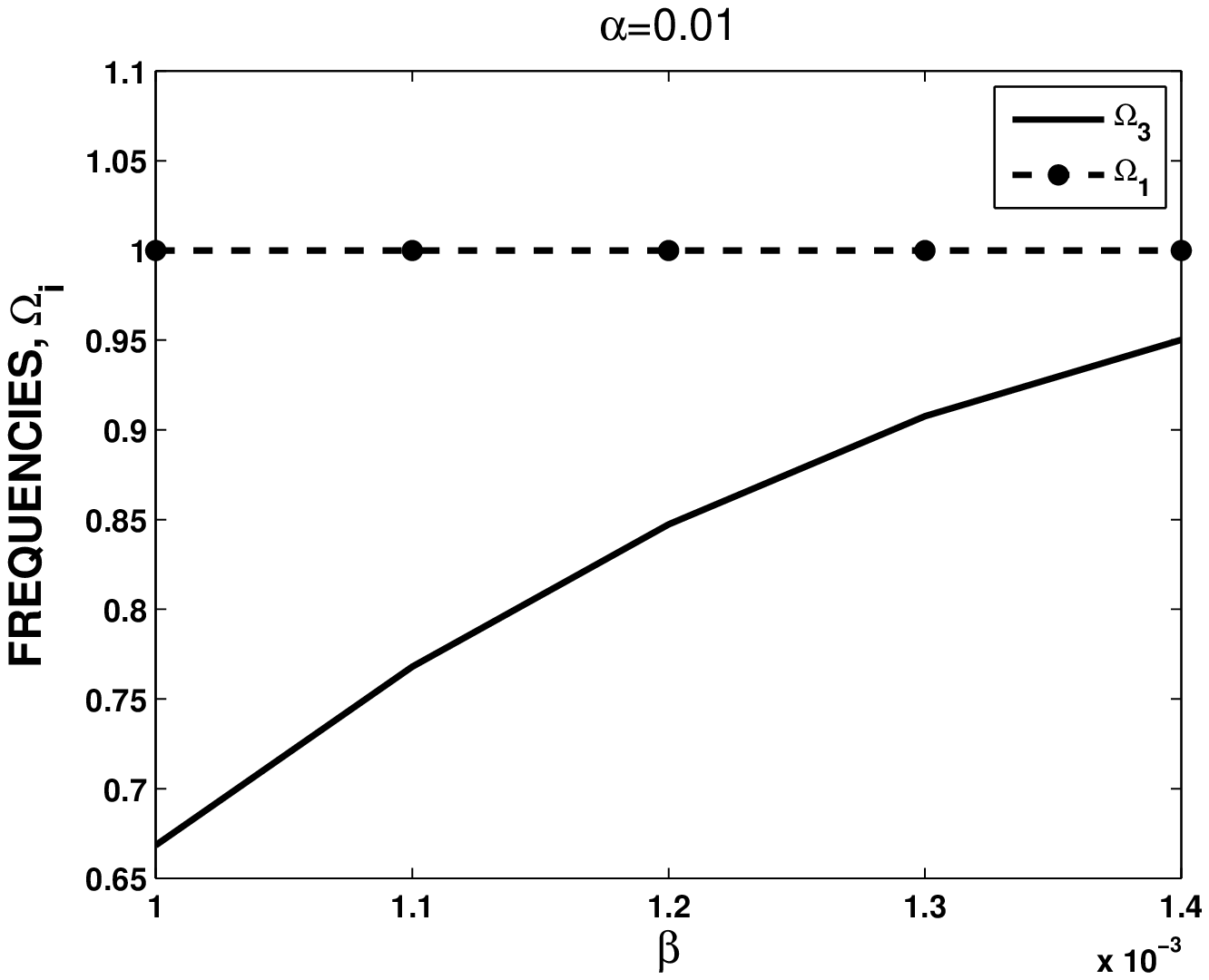,width=7.5cm,height=5cm,angle=0.0}}
\put(-10,0){\epsfig{file=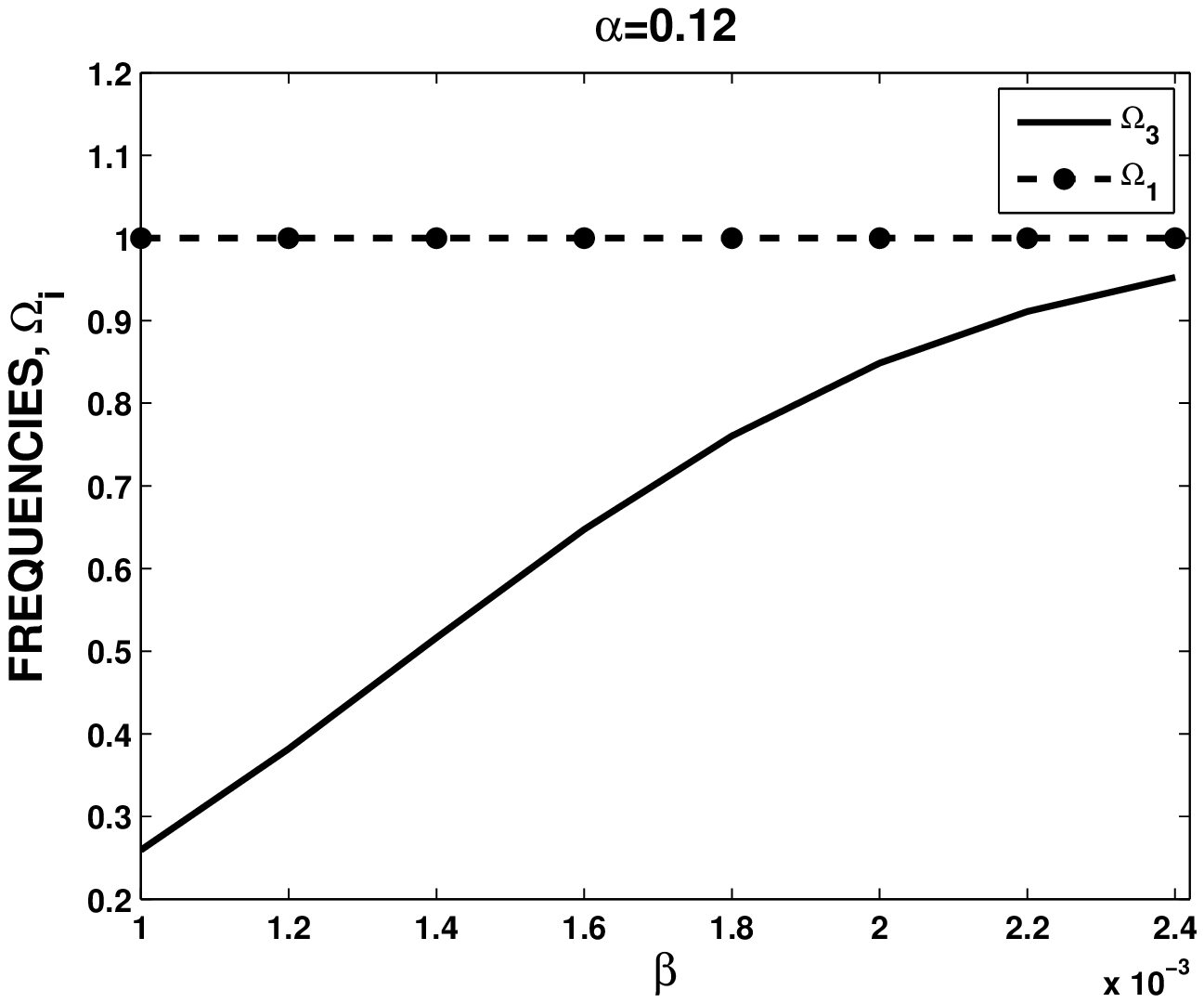,width=7.5cm,height=5cm,angle=0.0}}
\put(70,0){\epsfig{file=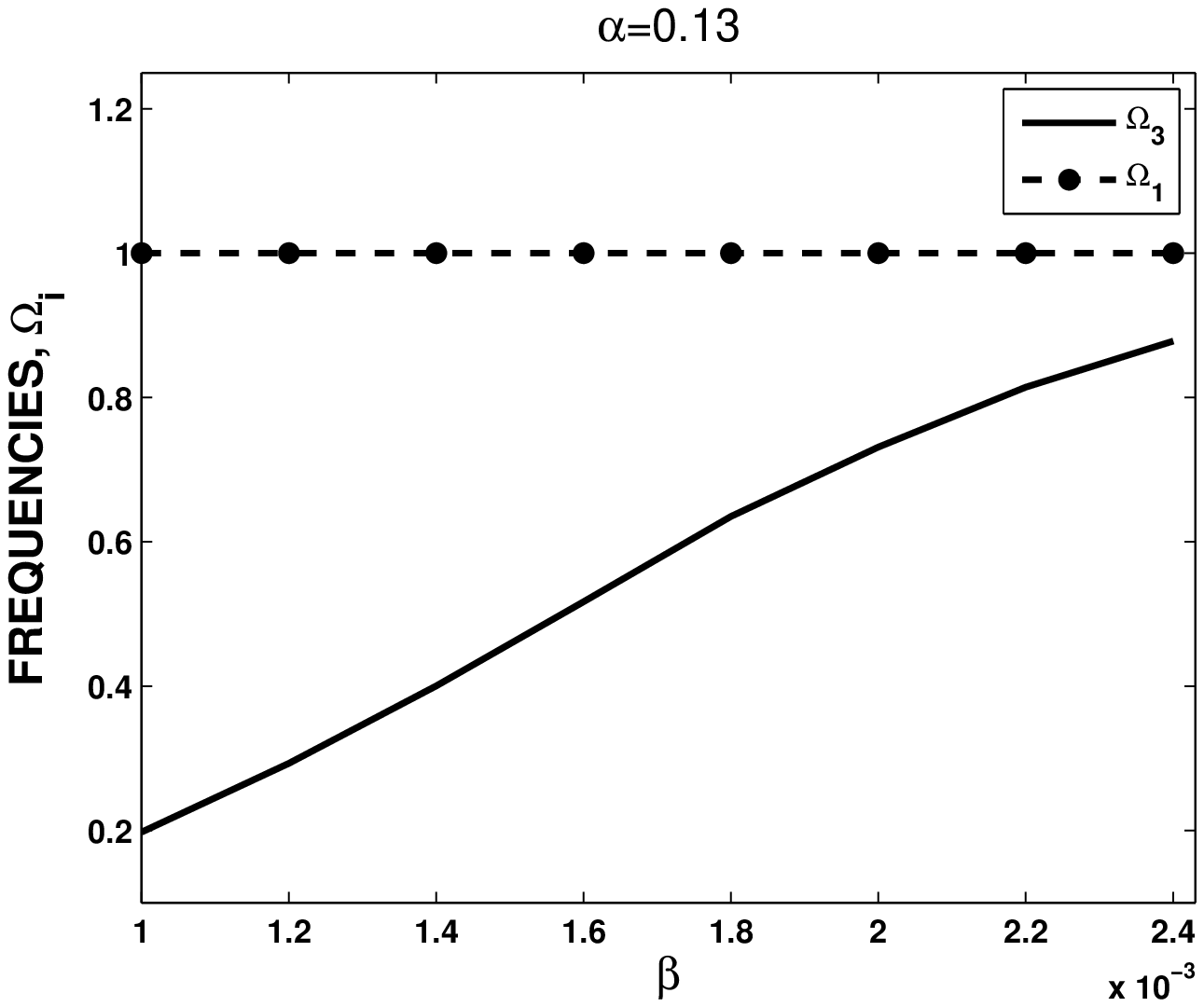,width=7.5cm,height=5cm,angle=0.0}}
\end{picture}
\caption[] {\footnotesize \it Frequency $\Omega_i$ versus the
parameter $\beta$ for  different values of $\alpha$ for the
noise-free self-sustained system. The nonlinear parameter reads
$\mu=0.1$.
 } \label{figurefreqb}
\end{center}
\end{minipage}
\end{figure}

\begin{figure}[htb]
\centering
\begin{minipage}{12cm}
\begin{center}
\begin{picture}(250,120)
\put(-10,50){\epsfig{file=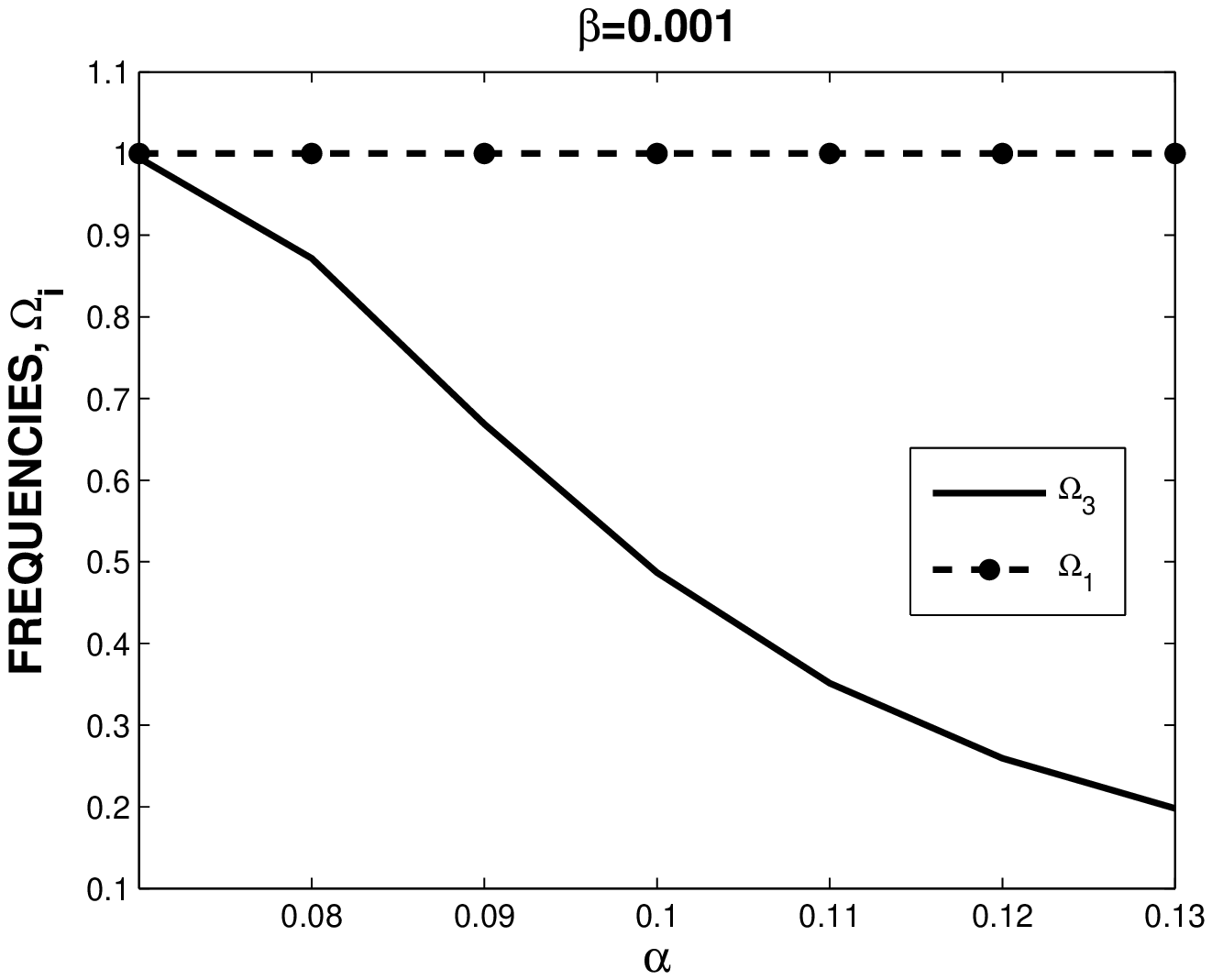,width=7.5cm,height=5cm,angle=0.0}}
\put(70,50){\epsfig{file=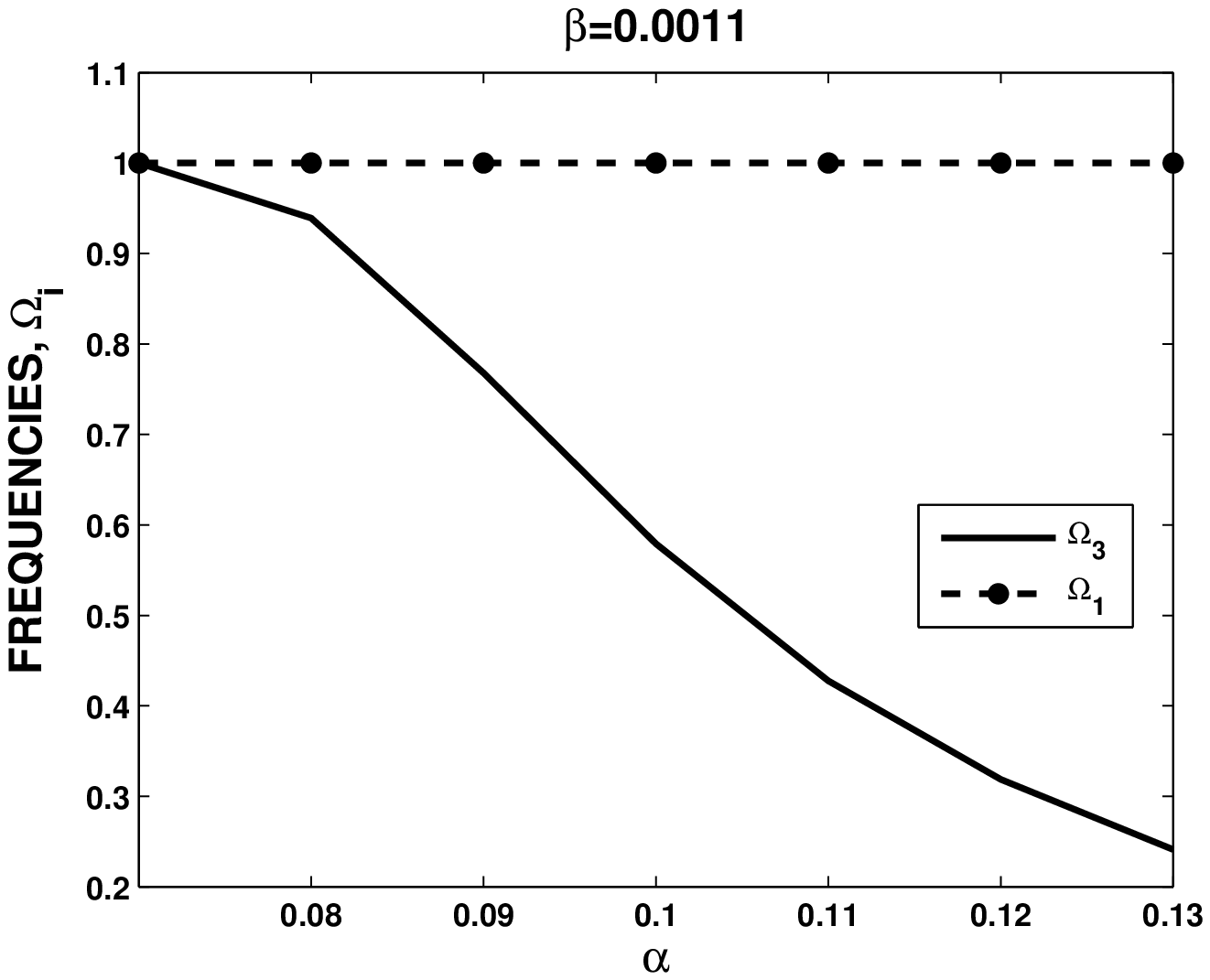,width=7.5cm,height=5cm,angle=0.0}}
\put(-10,0){\epsfig{file=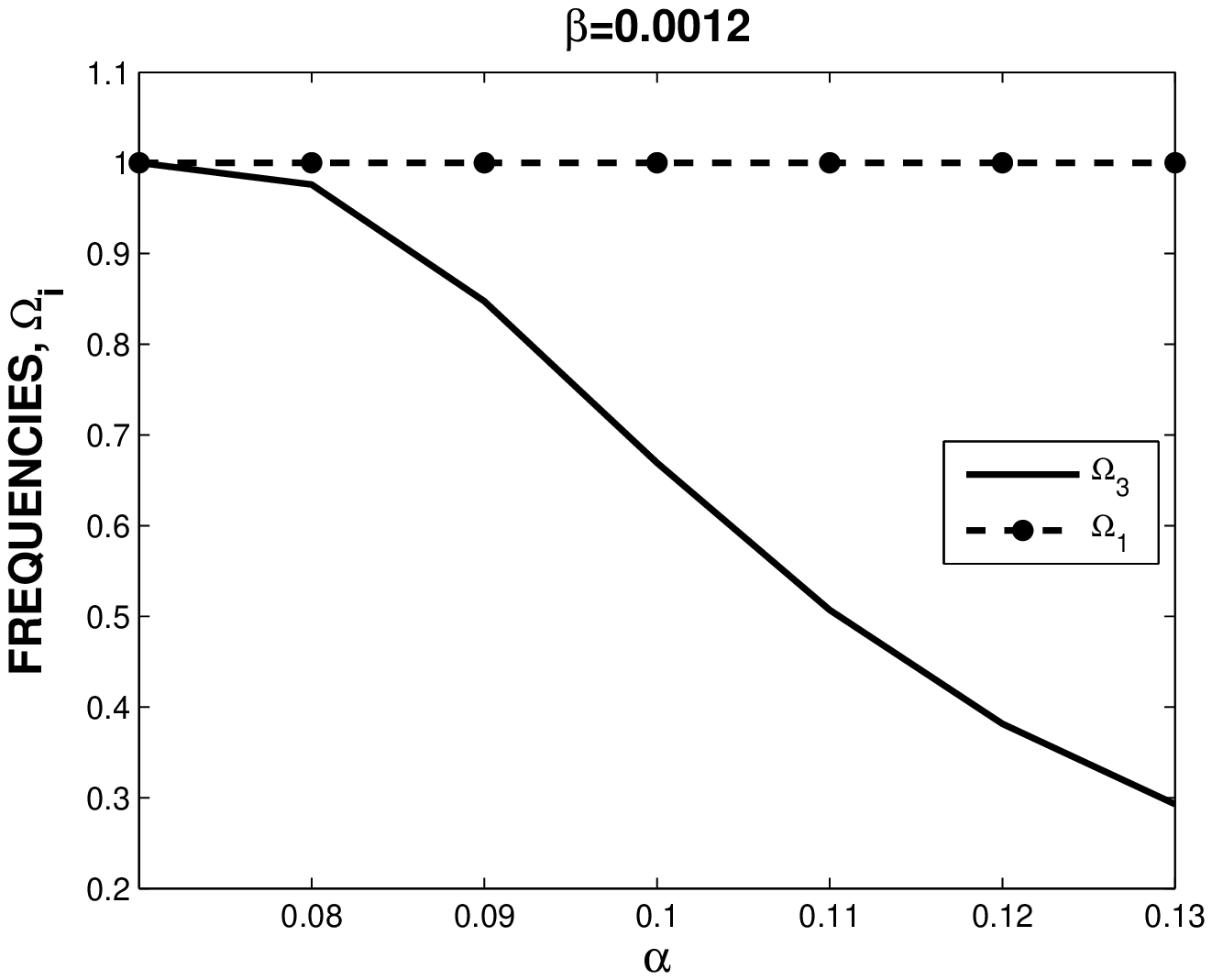,width=7.5cm,height=5cm,angle=0.0}}
\put(70,0){\epsfig{file=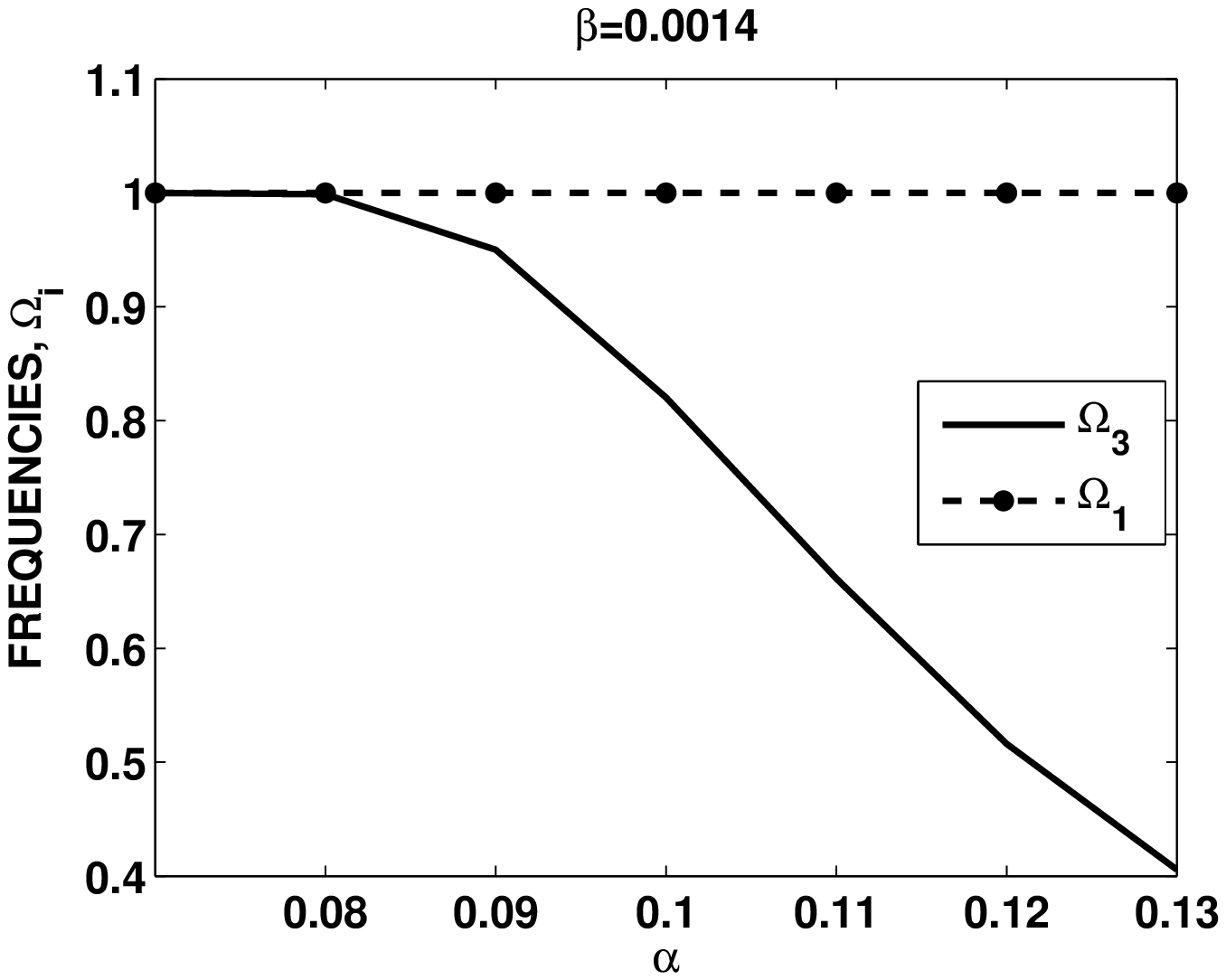,width=7.5cm,height=5cm,angle=0.0}}
\end{picture}
\caption[] {\footnotesize \it Frequency $\Omega_i$ versus the
parameter $\alpha$ for different values of $\beta$ for the
noise-free self-sustained system. The nonlinear parameter  reads
$\mu=0.1$.
 } \label{figurefreqa}
\end{center}
\end{minipage}
\end{figure}

\begin{figure}[htb]
\centering
\begin{minipage}{12cm}
\begin{center}
\begin{picture}(250,200)
\put(20,120){\epsfig{file=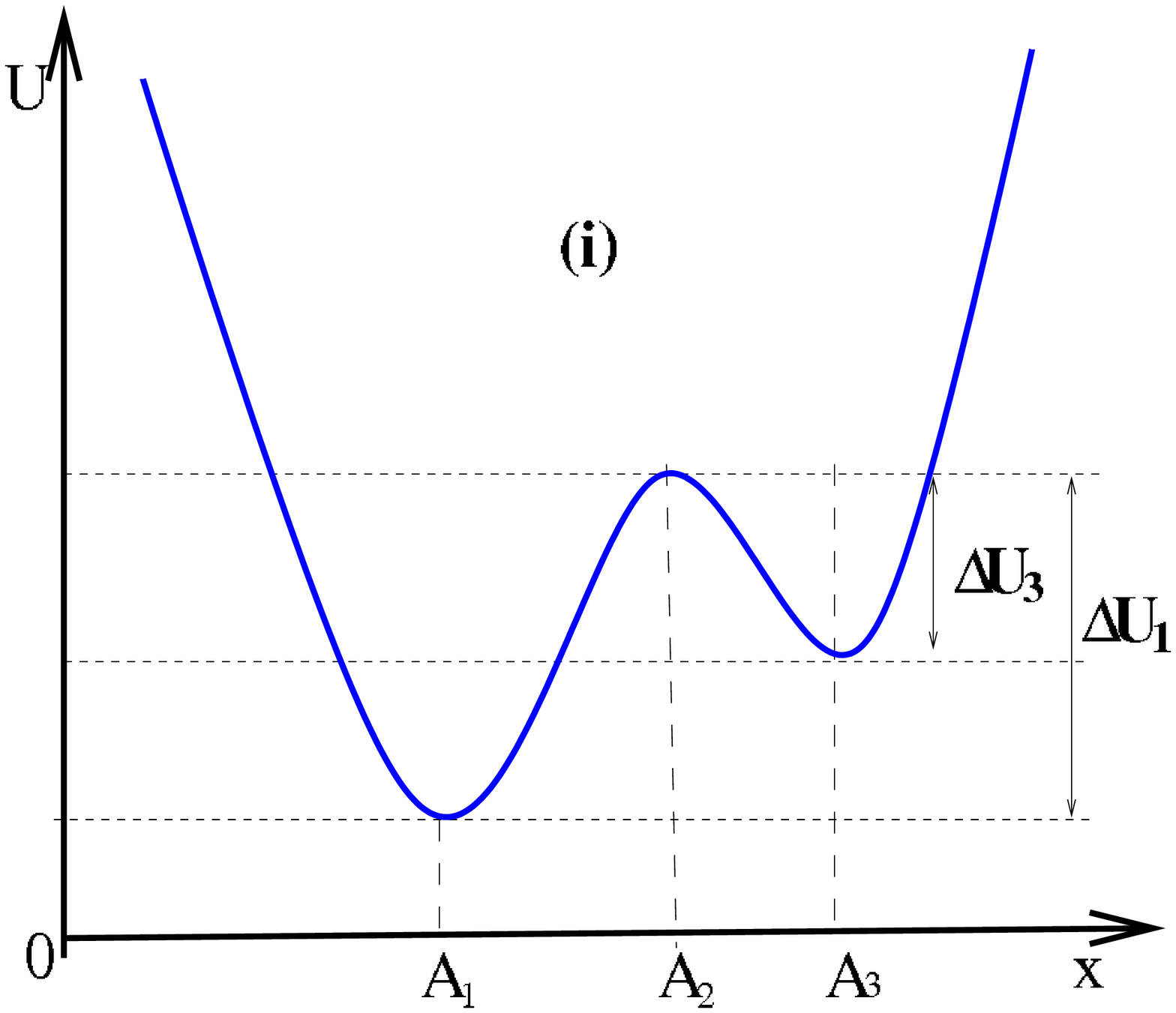,width=7cm,height=5cm,angle=0.0}}
\put(20,60){\epsfig{file=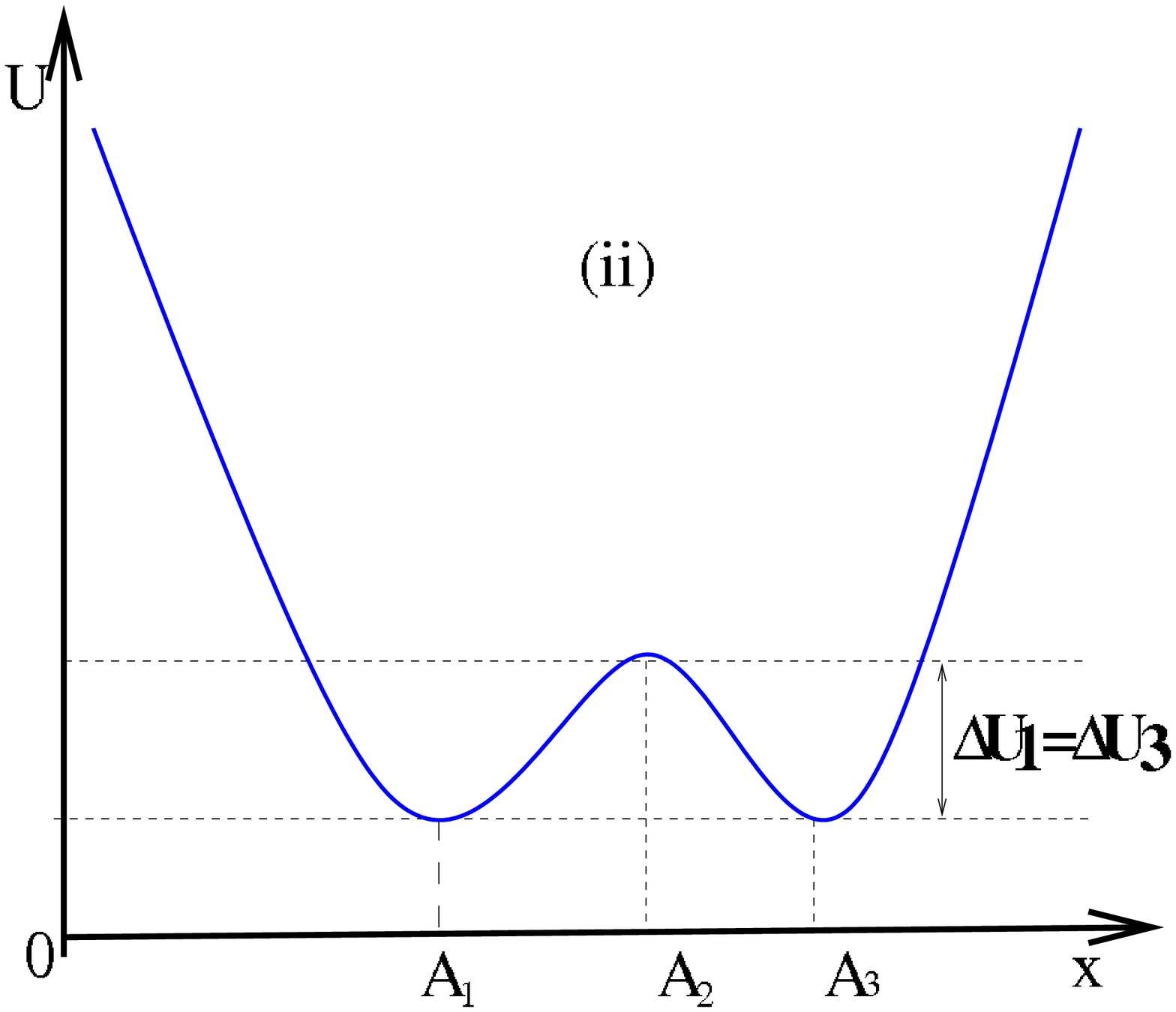,width=7cm,height=5cm,angle=0.0}}
\put(20,0){\epsfig{file=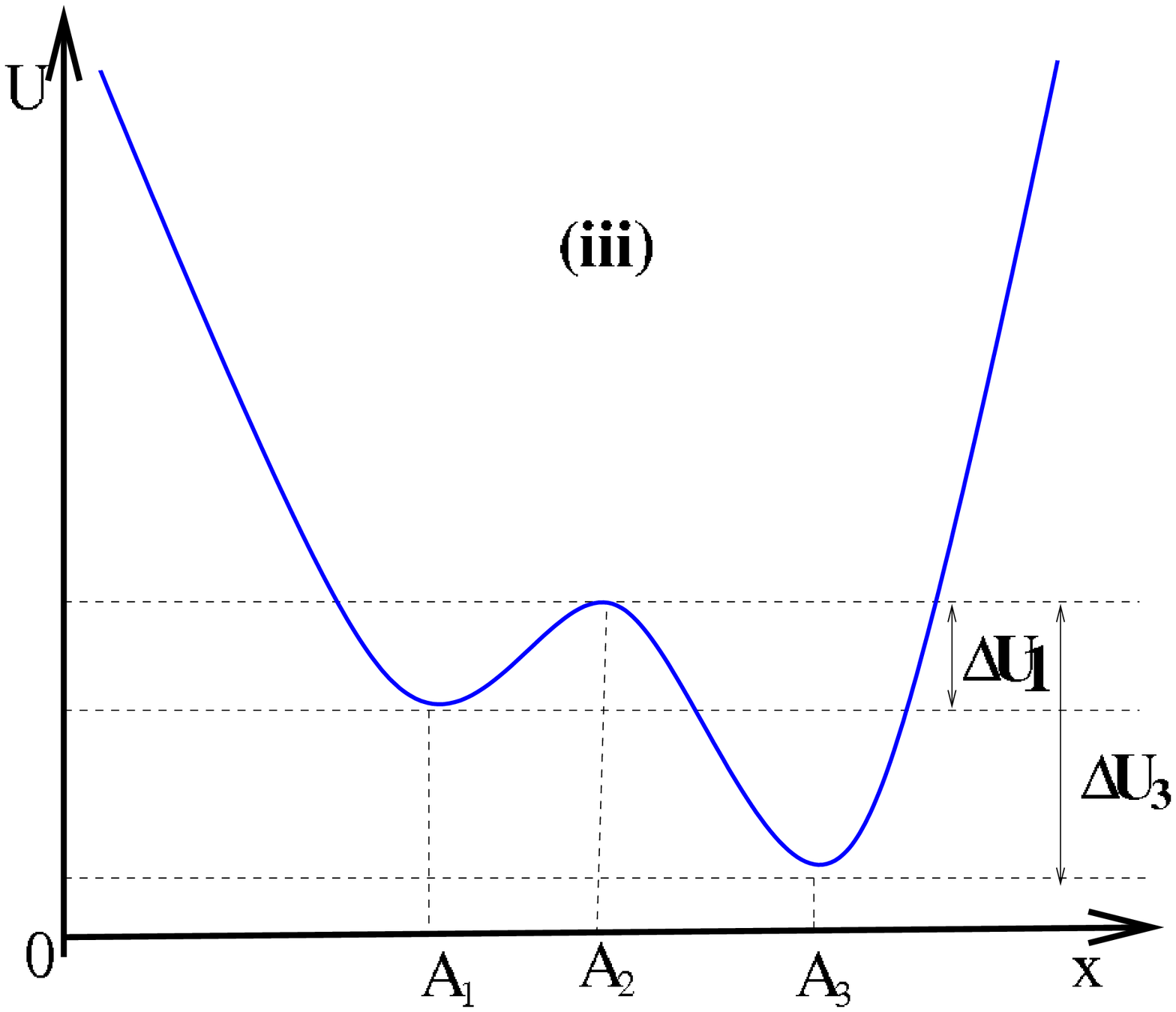,width=7cm,height=5cm,angle=0.0}}
\end{picture}
\caption[] {\footnotesize \it Sketch of the effective activation
energies $\Delta U_1$ and $\Delta U_3$ for the free-noise
self-sustained oscillator with multi-limit-cycles. We underline
that the barrier height has clear meaning as the slope of the
escape time \ref{eqtau}, while the effective potential $U$ is
qualitatively drawn only to help intuition.
 } \label{figureactivation}
\end{center}
\end{minipage}
\end{figure}

\begin{figure}[htb]
\centering
\begin{minipage}{12cm}
\begin{center}
\begin{picture}(250,150)
\put(0,80){\epsfig{file=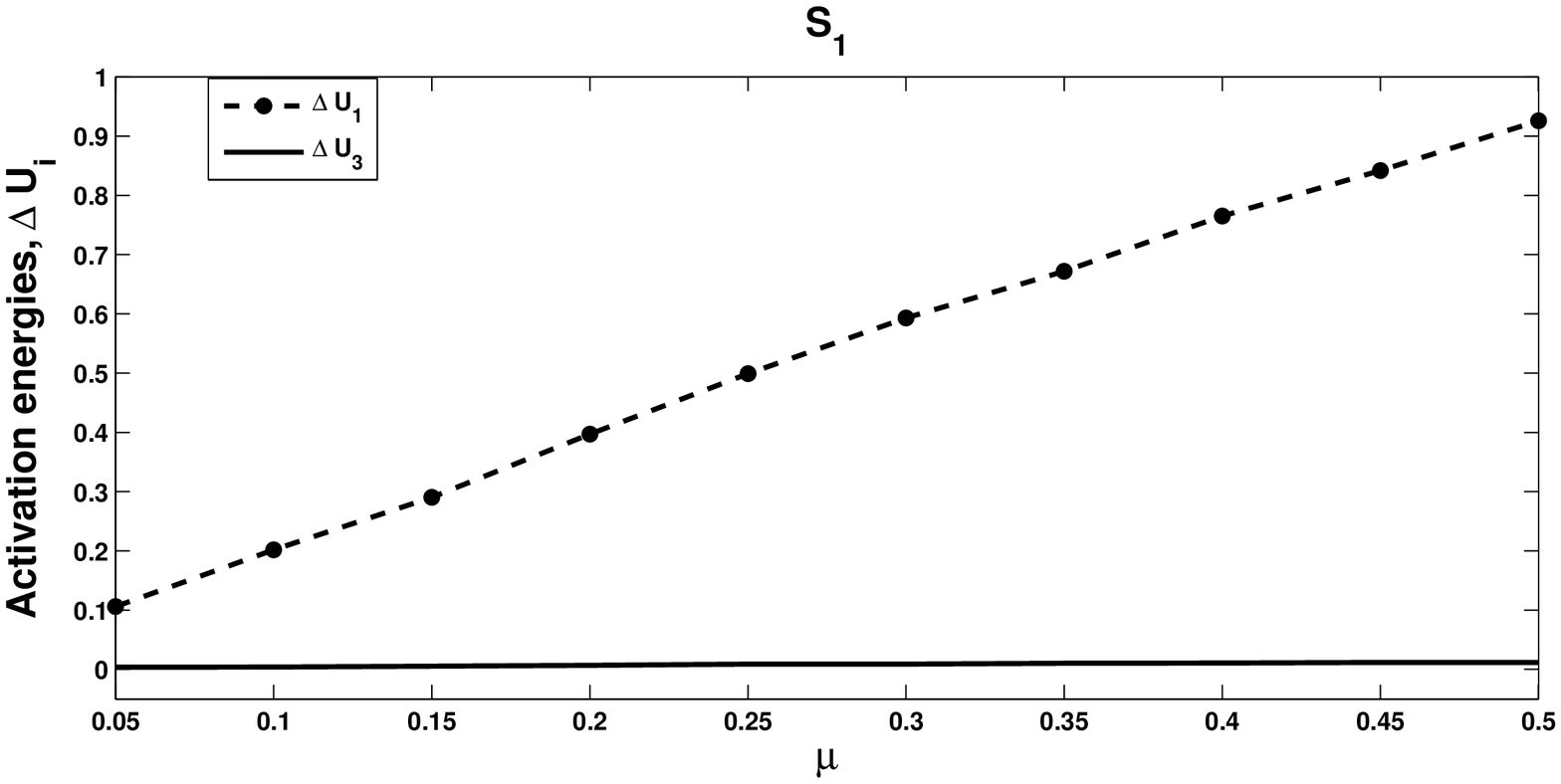,width=6cm,height=4cm,angle=0.0}}
\put(60,80){\epsfig{file=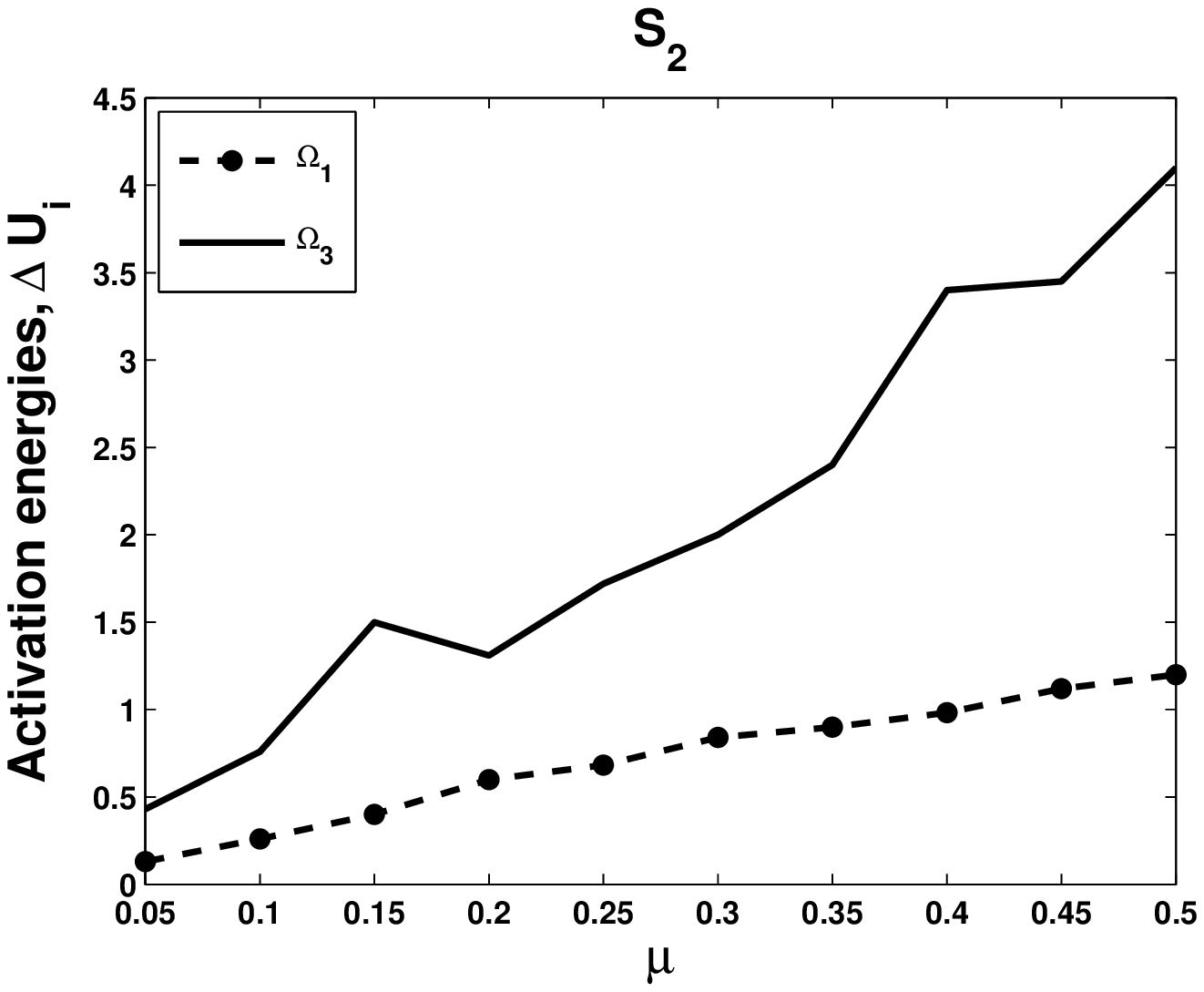,width=6cm,height=4cm,angle=0.0}}
\put(0,40){\epsfig{file=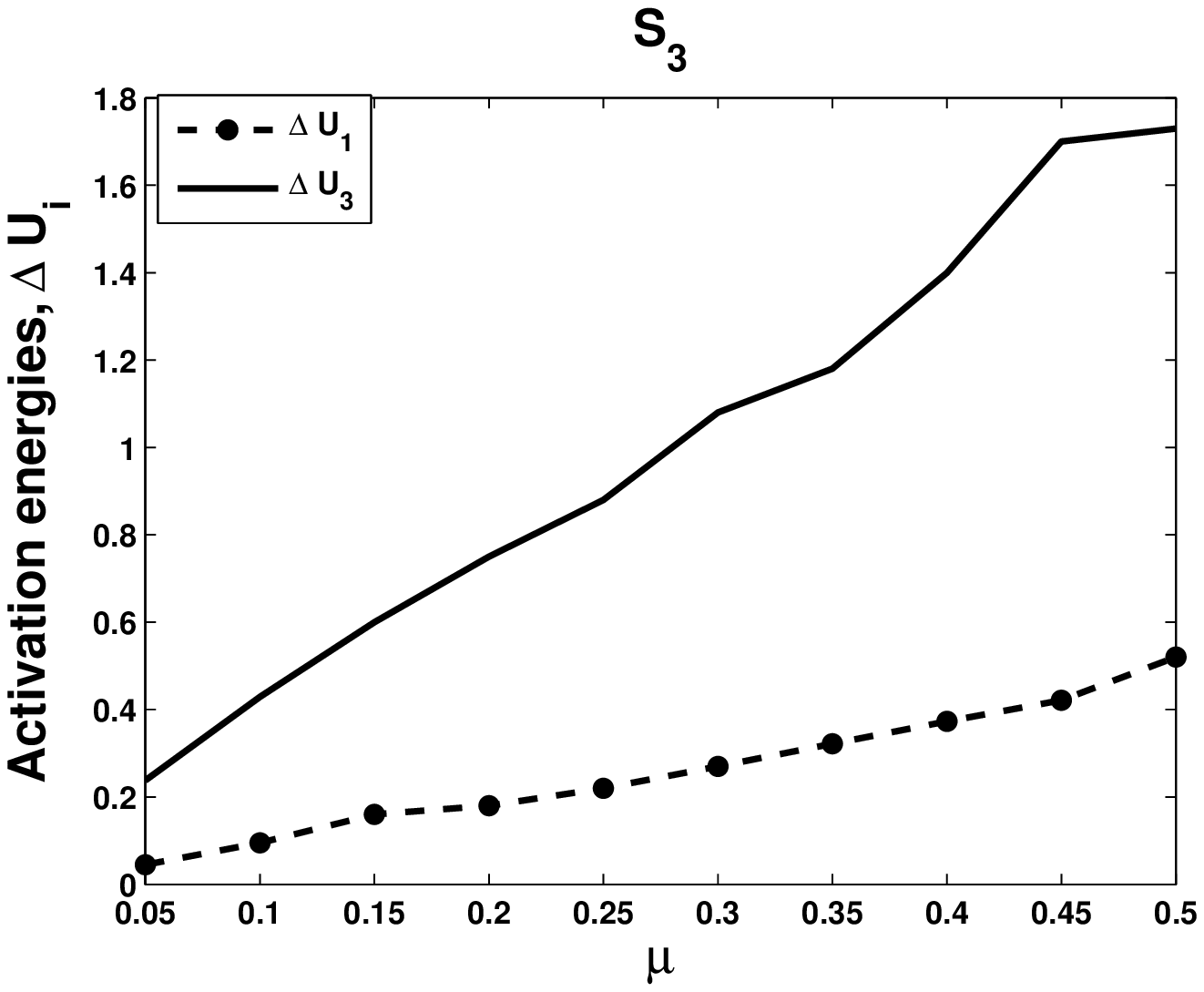,width=6cm,height=4cm,angle=0.0}}
\put(60,40){\epsfig{file=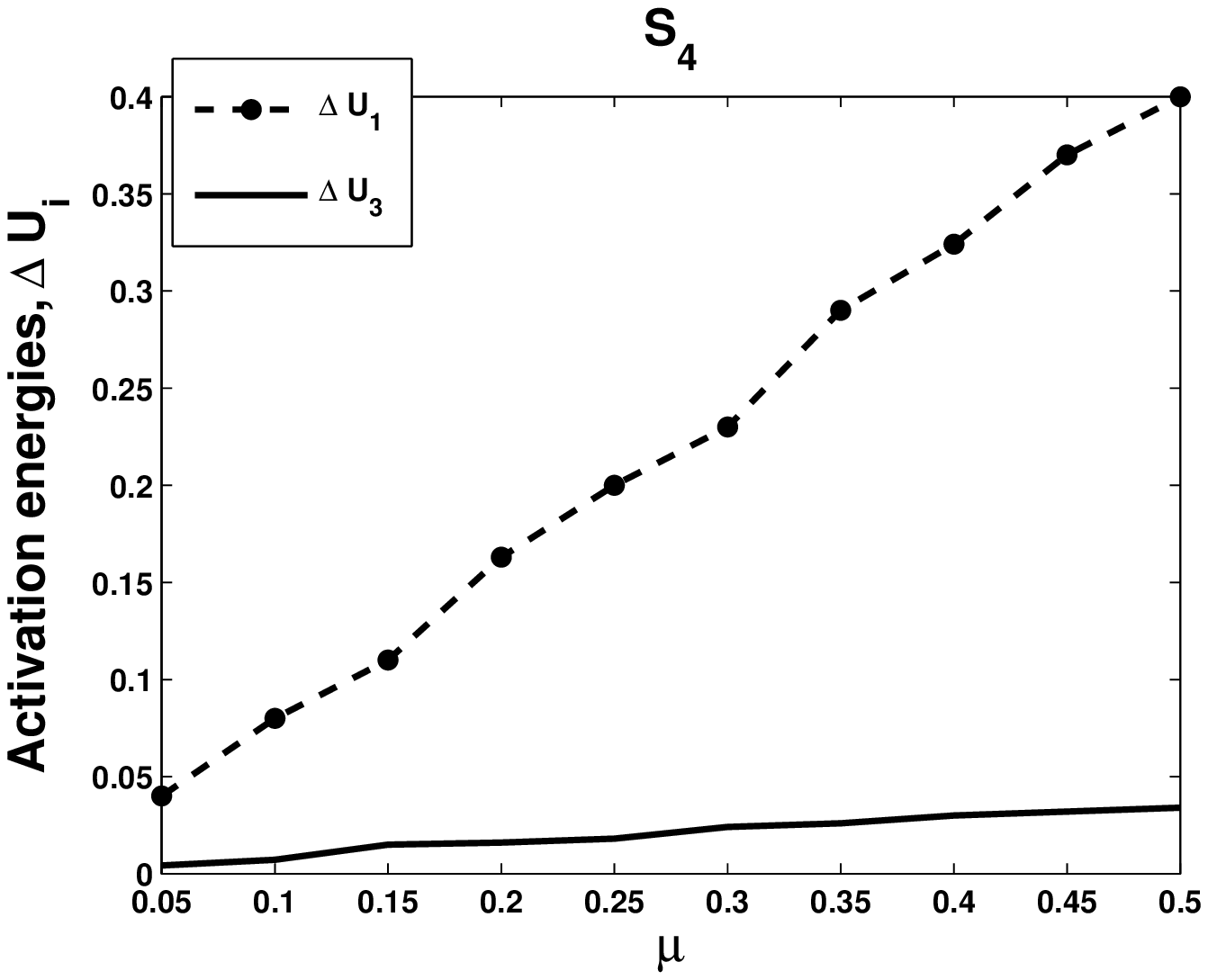,width=6cm,height=4cm,angle=0.0}}
\put(0,0){\epsfig{file=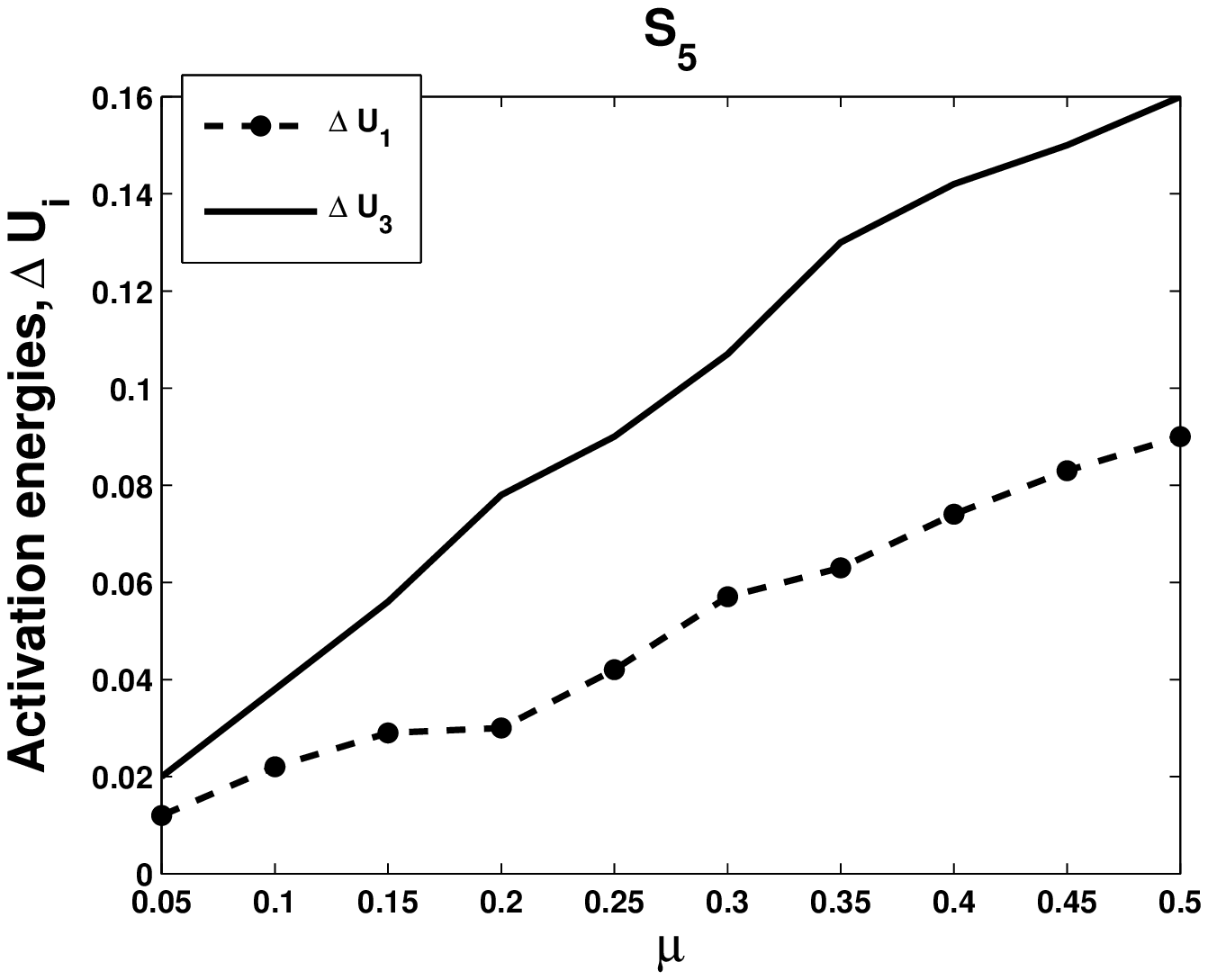,width=6cm,height=4cm,angle=0.0}}
\put(60,0){\epsfig{file=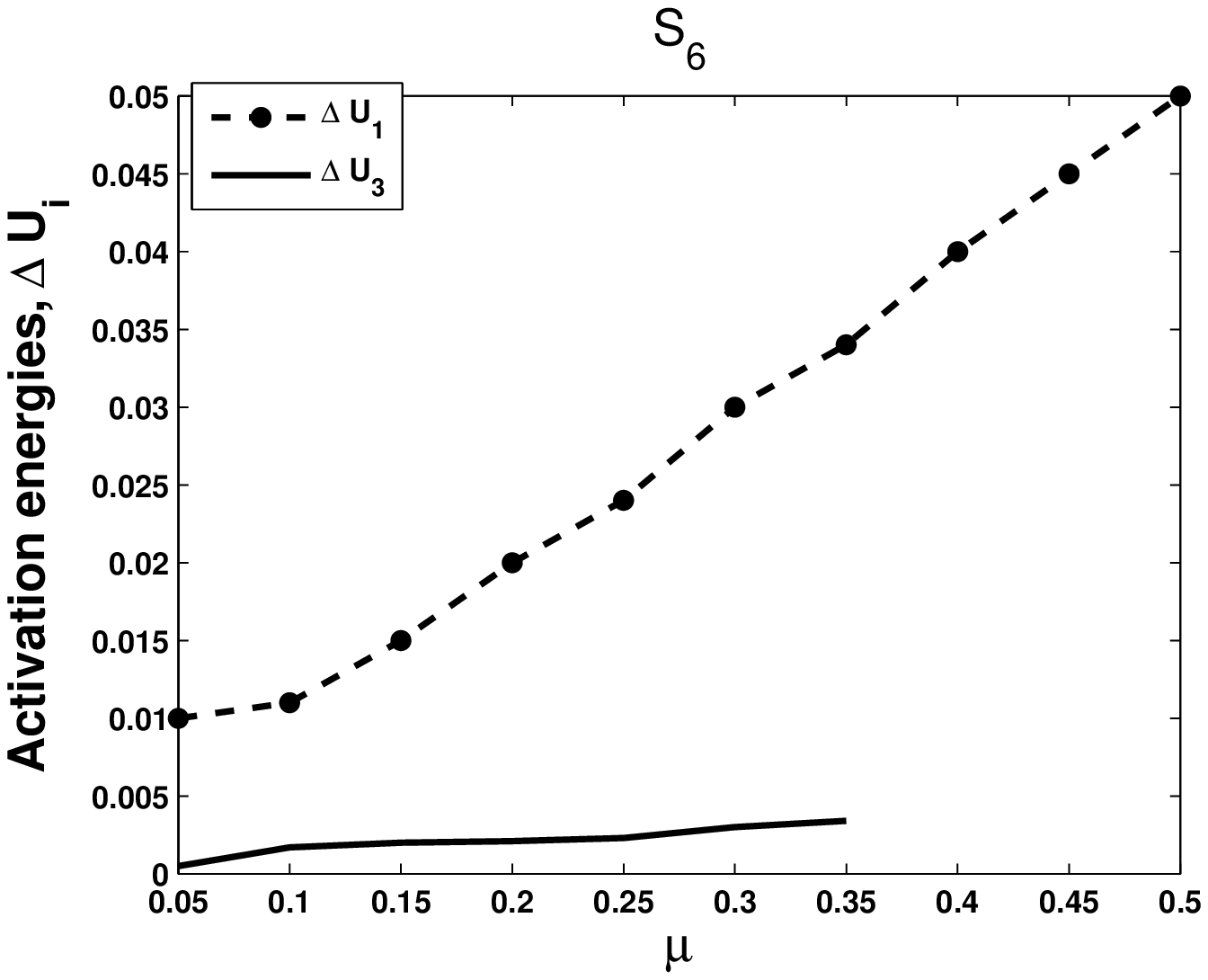,width=6cm,height=4cm,angle=0.0}}
\end{picture}
\caption[] {\footnotesize \it Effective activation energies versus
the coefficient $\mu$ with the set of parameters $S_i$. The thick
line corresponds to escape from the outer cycle $A_3$, while the
dashed line refers to escape from the inner cycle, $A_1$. The
parameters $\alpha$ and $\beta$ are the same as in Table $1$.
 } \label{figureactivationmu}
\end{center}
\end{minipage}
\end{figure}

\end{document}